\documentclass[12pt,a4paper]{article}

\usepackage{ifthen} 
\newboolean{pdflatex}
\setboolean{pdflatex}{true} 

\newboolean{articletitles}
\setboolean{articletitles}{true} 

\newboolean{uprightparticles}
\setboolean{uprightparticles}{false} 

\def\paperauthors{LHCb collaboration} 
\def\paperasciititle{Observation of the doubly-charmed-baryon decay Xicc++ -> Xic0 pi+ pi+ } 
\def\papertitle{Observation of the doubly-charmed-baryon decay ${\it \Xi}_{cc}^{++}\to{\it \Xi}_{c}^{0}\pi^{+}\pi^{+}$ } 
\def\paperkeywords{{Charm physics}, {Hadron-Hadron scattering (experiments)}, {Branching fraction},
{QCD}, {Flavor physics}} 
\def\papercopyright{\the\year\ CERN for the benefit of the LHCb collaboration} 
\def\paperlicence{CC BY 4.0 licence}
\def\paperlicenceurl{https://creativecommons.org/licenses/by/4.0/}

\newif\ifEnableSectionTOCLinks
\EnableSectionTOCLinksfalse 

\usepackage[top=1in, bottom=1.25in, left=1in, right=1in]{geometry}

\usepackage{microtype}
\usepackage{lineno}  
\usepackage{xspace} 
\usepackage{caption} 

\usepackage{graphicx}  
\usepackage{color}
\usepackage{colortbl}
\graphicspath{{./figs/}} 

\usepackage{amsmath} 
\usepackage{amssymb}
\usepackage{amsfonts}
\usepackage{upgreek} 

\newcommand*\patchAmsMathEnvironmentForLineno[1]{%
\expandafter\let\csname old#1\expandafter\endcsname\csname #1\endcsname
\expandafter\let\csname oldend#1\expandafter\endcsname\csname
end#1\endcsname
 \renewenvironment{#1}%
   {\linenomath\csname old#1\endcsname}%
   {\csname oldend#1\endcsname\endlinenomath}%
}
\newcommand*\patchBothAmsMathEnvironmentsForLineno[1]{%
  \patchAmsMathEnvironmentForLineno{#1}%
  \patchAmsMathEnvironmentForLineno{#1*}%
}
\AtBeginDocument{%
\patchBothAmsMathEnvironmentsForLineno{equation}%
\patchBothAmsMathEnvironmentsForLineno{align}%
\patchBothAmsMathEnvironmentsForLineno{flalign}%
\patchBothAmsMathEnvironmentsForLineno{alignat}%
\patchBothAmsMathEnvironmentsForLineno{gather}%
\patchBothAmsMathEnvironmentsForLineno{multline}%
\patchBothAmsMathEnvironmentsForLineno{eqnarray}%
}

\usepackage[pdftex,
            pdfauthor={\paperauthors},
            pdftitle={\paperasciititle},
            pdfkeywords={\paperkeywords}]{hyperref}
\usepackage{hyperxmp}
\hypersetup{
    pdfcopyright={Copyright (C) \papercopyright},
    pdflicenseurl={\paperlicenceurl}
}

\usepackage[colorinlistoftodos,textsize=scriptsize]{todonotes}

\usepackage[bottom,flushmargin,hang,multiple]{footmisc}

\usepackage[all]{hypcap} 

\usepackage{xspace}
\usepackage{upgreek}

\def\lhcb   {\mbox{LHCb}\xspace}

\def\MagUp {\mbox{\em Mag\kern -0.05em Up}\xspace}

\ifthenelse{\boolean{uprightparticles}}%
{
 
 \def\Pgamma      {\ensuremath{\upgamma}\xspace}

 \def\Pmu         {\ensuremath{\upmu}\xspace}

 \def\Ppi         {\ensuremath{\uppi}\xspace}

 \def\Ppsi        {\ensuremath{\uppsi}\xspace}

 \def\PDelta      {\ensuremath{\Delta}\xspace}
 \def\PXi         {\ensuremath{\Xi}\xspace}
 \def\PLambda     {\ensuremath{\Lambda}\xspace}
 \def\PSigma      {\ensuremath{\Sigma}\xspace}
 \def\POmega      {\ensuremath{\Omega}\xspace}
 \def\PUpsilon    {\ensuremath{\Upsilon}\xspace}
 \let\oldPi\Pi
 \def\PPi         {\ensuremath{\oldPi}\xspace}

 \def\PB      {\ensuremath{\mathrm{B}}\xspace}
 \def\PD      {\ensuremath{\mathrm{D}}\xspace}
 \def\PJ      {\ensuremath{\mathrm{J}}\xspace}
 \def\PK      {\ensuremath{\mathrm{K}}\xspace}
 \def\Pb      {\ensuremath{\mathrm{b}}\xspace}
 \def\Pc      {\ensuremath{\mathrm{c}}\xspace}
 \def\Pd      {\ensuremath{\mathrm{d}}\xspace}

 \def\Pp      {\ensuremath{\mathrm{p}}\xspace}

 \def\Ps      {\ensuremath{\mathrm{s}}\xspace}
 
 \def\Pu      {\ensuremath{\mathrm{u}}\xspace}
 \def\thebaroffset{0.0em}
}
{
 
 \def\Pgamma      {\ensuremath{\gamma}\xspace}

 \def\Pmu         {\ensuremath{\mu}\xspace}

 \def\Ppi         {\ensuremath{\pi}\xspace}

 \def\Ppsi        {\ensuremath{\psi}\xspace}
 
 \mathchardef\PDelta="7101
 \mathchardef\PXi="7104
 \mathchardef\PLambda="7103
 \mathchardef\PSigma="7106
 \mathchardef\POmega="710A
 \mathchardef\PUpsilon="7107
 \mathchardef\PPi="7105
 \def\PB      {\ensuremath{B}\xspace}
 \def\PD      {\ensuremath{D}\xspace}
 \def\PJ      {\ensuremath{J}\xspace}
 \def\PK      {\ensuremath{K}\xspace}
 \def\Pb      {\ensuremath{b}\xspace}
 \def\Pc      {\ensuremath{c}\xspace}
 \def\Pd      {\ensuremath{d}\xspace}

 \def\Pp      {\ensuremath{p}\xspace}

 \def\Ps      {\ensuremath{s}\xspace}
 
 \def\Pu      {\ensuremath{u}\xspace}
 \def\thebaroffset{0.18em}
}
\newcommand{\offsetoverline}[2][\thebaroffset]{\kern #1\overline{\kern -#1 #2}}%

\makeatletter
\ifcase \@ptsize \relax
  \newcommand{\miniscule}{\@setfontsize\miniscule{4}{5}}
\or
  \newcommand{\miniscule}{\@setfontsize\miniscule{5}{6}}
\or
  \newcommand{\miniscule}{\@setfontsize\miniscule{5}{6}}
\fi
\makeatother

\DeclareRobustCommand{\optbar}[1]{\shortstack{{\miniscule (\rule[.5ex]{1.25em}{.18mm})}
  \\ [-.7ex] $#1$}}

\def\mumu       {{\ensuremath{\Pmu^+\Pmu^-}}\xspace}

\def\uquark    {{\ensuremath{\Pu}}\xspace}

\def\dquark    {{\ensuremath{\Pd}}\xspace}

\def\squark    {{\ensuremath{\Ps}}\xspace}

\def\cquark    {{\ensuremath{\Pc}}\xspace}

\def\bquark    {{\ensuremath{\Pb}}\xspace}

\def\pion   {{\ensuremath{\Ppi}}\xspace}

\def\pip    {{\ensuremath{\pion^+}}\xspace}
\def\pim    {{\ensuremath{\pion^-}}\xspace}

\def\kaon    {{\ensuremath{\PK}}\xspace}
\def\Kbar    {{\ensuremath{\offsetoverline{\PK}}}\xspace}

\def\KorKbar {\kern \thebaroffset\optbar{\kern -\thebaroffset \PK}{}\xspace}

\def\Kp      {{\ensuremath{\kaon^+}}\xspace}
\def\Km      {{\ensuremath{\kaon^-}}\xspace}

\def\KS      {{\ensuremath{\kaon^0_{\mathrm{S}}}}\xspace}

\def\Kstarb  {{\ensuremath{\Kbar{}^*}}\xspace}

\def\D       {{\ensuremath{\PD}}\xspace}

\def\DorDbar {\kern \thebaroffset\optbar{\kern -\thebaroffset \PD}\xspace}

\def\Dp      {{\ensuremath{\D^+}}\xspace}
\def\Dm      {{\ensuremath{\D^-}}\xspace}

\def\DpDm    {\ensuremath{\Dp {\kern -0.16em \Dm}}\xspace}

\def\B       {{\ensuremath{\PB}}\xspace}

\def\BorBbar {\kern \thebaroffset\optbar{\kern -\thebaroffset \PB}\xspace}

\def\Bd      {{\ensuremath{\B^0}}\xspace}

\def\BdorBdbar {\kern \thebaroffset\optbar{\kern -\thebaroffset \Bd}\xspace}
\def\Bu      {{\ensuremath{\B^+}}\xspace}

\def\Bs      {{\ensuremath{\B^0_\squark}}\xspace}

\def\BsorBsbar {\kern \thebaroffset\optbar{\kern -\thebaroffset \Bs}\xspace}

\def\jpsi     {{\ensuremath{{\PJ\mskip -3mu/\mskip -2mu\Ppsi}}}\xspace}

\def\Y#1S{\ensuremath{\PUpsilon{(#1S)}}\xspace}

\def\proton      {{\ensuremath{\Pp}}\xspace}

\def\Lz          {{\ensuremath{\PLambda}}\xspace}

\def\LorLbar     {\kern \thebaroffset\optbar{\kern -\thebaroffset \PLambda}\xspace}

\def\Xires       {{\ensuremath{\PXi}}\xspace}

\def\Omegares    {{\ensuremath{\POmega}}\xspace}

\def\Lc          {{\ensuremath{\Lz^+_\cquark}}\xspace}

\def\Xicz        {{\ensuremath{\Xires^0_\cquark}}\xspace}
\def\Xicp        {{\ensuremath{\Xires^+_\cquark}}\xspace}

\def\Xiccp       {{\ensuremath{\Xires^+_{\cquark\cquark}}}\xspace}
\def\Xiccpp      {{\ensuremath{\Xires^{++}_{\cquark\cquark}}}\xspace}

\def\Omegacc     {{\ensuremath{\Omegares^+_{\cquark\cquark}}}\xspace}

\def\Xibm         {{\ensuremath{\Xires^-_\bquark}}\xspace}

\def\BF         {{\ensuremath{\mathcal{B}}}\xspace}

\newcommand{\decay}[2]{\mbox{\ensuremath{#1\!\to #2}}\xspace}

\def\to                 {\ensuremath{\rightarrow}\xspace}

\def\AT#1     {\ensuremath{A_{\mathrm{T}}^{#1}}\xspace}           

\def\C#1      {\ensuremath{\mathcal{C}_{#1}}\xspace}                       
\def\Cp#1     {\ensuremath{\mathcal{C}_{#1}^{'}}\xspace}                    
\def\Ceff#1   {\ensuremath{\mathcal{C}_{#1}^{\mathrm{(eff)}}}\xspace}        
\def\Cpeff#1  {\ensuremath{\mathcal{C}_{#1}^{'\mathrm{(eff)}}}\xspace}       
\def\Ope#1    {\ensuremath{\mathcal{O}_{#1}}\xspace}                       
\def\Opep#1   {\ensuremath{\mathcal{O}_{#1}^{'}}\xspace}                    

\newcommand{\nospaceunit}[1]{\ensuremath{\text{#1}}}
\newcommand{\aunit}[1]{\ensuremath{\text{\,#1}}}

\newcommand{\tev}{\aunit{Te\kern -0.1em V}\xspace}
\newcommand{\gev}{\aunit{Ge\kern -0.1em V}\xspace}
\newcommand{\mev}{\aunit{Me\kern -0.1em V}\xspace}
\newcommand{\kev}{\aunit{ke\kern -0.1em V}\xspace}
\newcommand{\ev}{\aunit{e\kern -0.1em V}\xspace}

\newcommand{\mevc}{\ensuremath{\aunit{Me\kern -0.1em V\!/}c}\xspace}
\newcommand{\gevc}{\ensuremath{\aunit{Ge\kern -0.1em V\!/}c}\xspace}
\newcommand{\mevcc}{\ensuremath{\aunit{Me\kern -0.1em V\!/}c^2}\xspace}
\newcommand{\gevcc}{\ensuremath{\aunit{Ge\kern -0.1em V\!/}c^2}\xspace}

\def\mum  {\ensuremath{\,\upmu\nospaceunit{m}}\xspace}

\def\fb   {\ensuremath{\aunit{fb}}\xspace}
\def\invfb   {\ensuremath{\fb^{-1}}\xspace}

\def\ps   {\ensuremath{\aunit{ps}}\xspace}

\newcommand{\stat}{\aunit{(stat)}\xspace}
\newcommand{\syst}{\aunit{(syst)}\xspace}

\newcommand{\chisq}{\ensuremath{\chi^2}\xspace}

\newcommand{\chisqip}{\ensuremath{\chi^2_{\text{IP}}}\xspace}

\def\gsim{{~\raise.15em\hbox{$>$}\kern-.85em
          \lower.35em\hbox{$\sim$}~}\xspace}
\def\lsim{{~\raise.15em\hbox{$<$}\kern-.85em
          \lower.35em\hbox{$\sim$}~}\xspace}

\def\sPlot{\mbox{\em sPlot}\xspace}

\def\pt         {\ensuremath{p_{\mathrm{T}}}\xspace}

\def\ptot       {\ensuremath{p}\xspace}

\def\evtgen     {\mbox{\textsc{EvtGen}}\xspace}

\def\geant      {\mbox{\textsc{Geant4}}\xspace}

\def\photos     {\mbox{\textsc{Photos}}\xspace}

\def\pythia     {\mbox{\textsc{Pythia}}\xspace}

\def\tell1  {TELL1\xspace}
\def\ukl1   {UKL1\xspace}

\newcommand{\ie}{\mbox{\itshape i.e.}\xspace}

\newcommand{\lhcborcid}[1]{\href{https://orcid.org/#1}{\hspace*{0.1em}\raisebox{-0.45ex}{\includegraphics[width=1em]{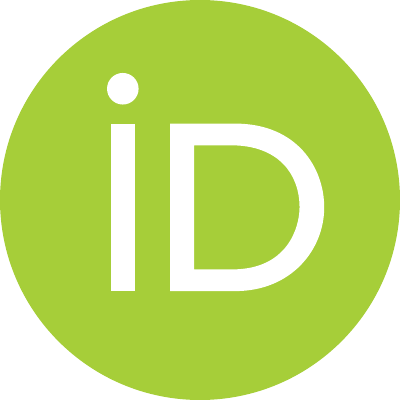}}}}

\hypersetup{
  colorlinks   = true, 
  urlcolor     = blue, 
  linkcolor    = blue, 
  citecolor    = red   
}

\ifEnableSectionTOCLinks
    \usepackage[explicit]{titlesec} 
    
    \let\oldcontentsline\contentsline
    \renewcommand

    \titleformat{\section}{\normalfont\Large\bf}{\hyperlink{tocsection.\thesection}{{\thesection} \parbox[t]{\dimexpr\textwidth-1pc}{#1}}}{1pc}{}

    \titleformat{\subsection}{\normalfont\bf}{\hyperlink{tocsubsection.\thesubsection}{{\thesubsection} \parbox[t]{\dimexpr\textwidth-1pc}{#1}}}{1pc}{}

    \titleformat{name=\section,numberless}[display]{}{}{0pt}{\normalfont\Huge\bfseries #1}
\fi

\usepackage{cite} 
\usepackage{mciteplus}

\usepackage{longtable} 
\usepackage{multirow}
\usepackage{verbatim}
\usepackage{overpic}
\usepackage{ifthen} 
\usepackage{booktabs}

\def\genxicc     {\mbox{\textsc{GenXicc}}\xspace}

\def\Xicppr        {{\ensuremath{\Xires^{\prime+}_\cquark}}\xspace}
\def\Xicporpr        {{\ensuremath{\Xires^{(\prime)+}_\cquark}}\xspace}
\begin{document}

\renewcommand{\thefootnote}{\fnsymbol{footnote}}
\setcounter{footnote}{1}


\begin{titlepage}
\pagenumbering{roman}

\vspace*{-1.5cm}
\centerline{\large EUROPEAN ORGANIZATION FOR NUCLEAR RESEARCH (CERN)}
\vspace*{1.5cm}
\noindent
\begin{tabular*}{\linewidth}{lc@{\extracolsep{\fill}}r@{\extracolsep{0pt}}}
\ifthenelse{\boolean{pdflatex}}
{\vspace*{-1.5cm}\mbox{\!\!\!\includegraphics[width=.14\textwidth]{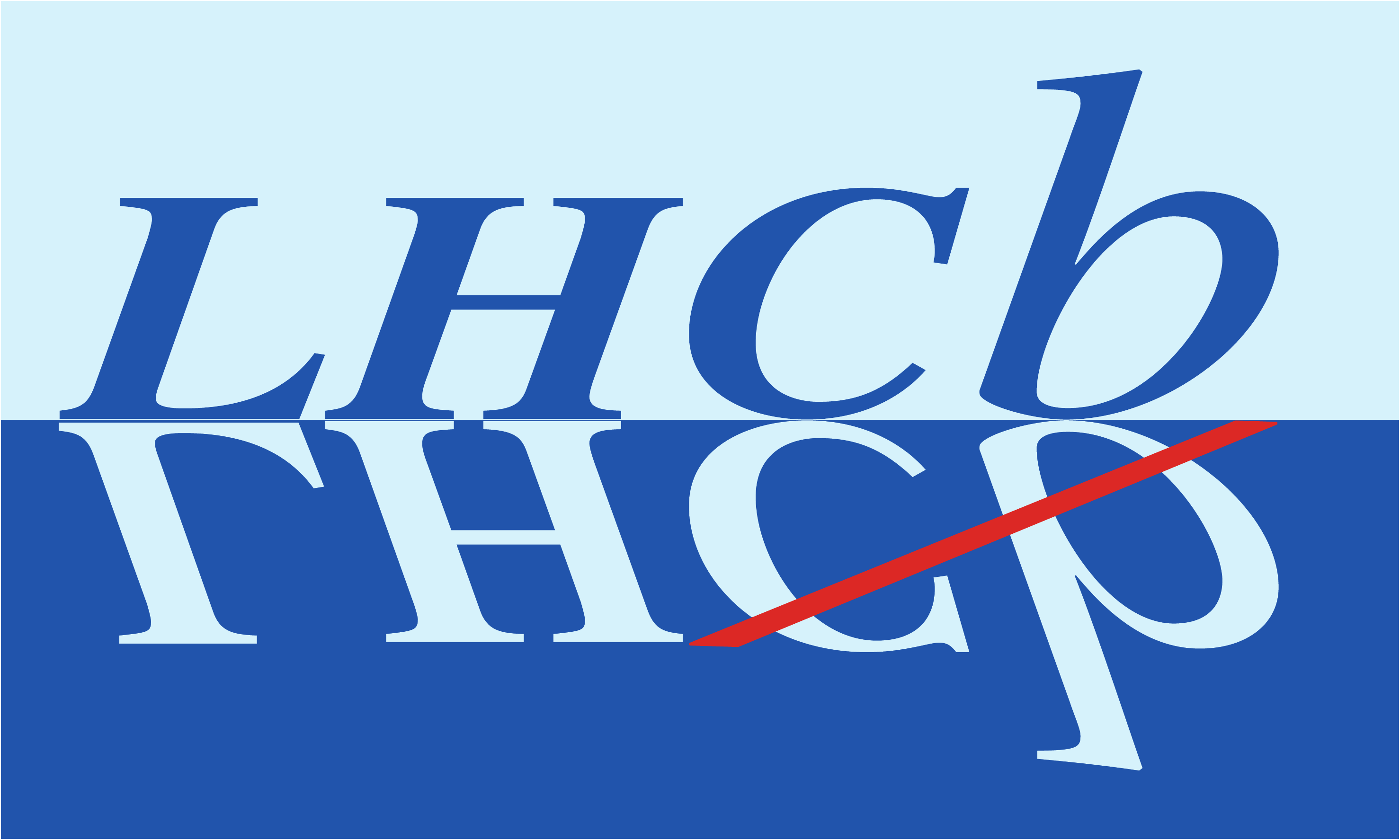}} & &}%
{\vspace*{-1.2cm}\mbox{\!\!\!\includegraphics[width=.12\textwidth]{figs/lhcb-logo.pdf}} & &}%
\\
 & & CERN-EP-2025-047 \\  
 & & LHCb-PAPER-2024-053 \\  
 & & October 30, 2025 \\ 
 & & \\
\end{tabular*}

\vspace*{2.0cm}

{\normalfont\bfseries\boldmath\huge
\begin{center}
  \papertitle 
\end{center}
}

\vspace*{2.0cm}

\begin{center}
\paperauthors\footnote{Authors are listed at the end of this paper.}
\end{center}

\vspace{\fill}

\begin{abstract}
  \noindent
A search for the doubly-charmed-baryon decay ${\it \Xi}_{cc}^{++}\to{\it \Xi}_{c}^{0}\pi^{+}\pi^{+}$ is performed using proton-proton collision data collected by the LHCb experiment at a centre-of-mass energy of $13\text{\,Te\kern -0.1em V}$ and corresponding to an integrated luminosity of $5.4\text{\,fb}^{-1}$. A significant structure consistent with the ${\it \Xi}_{cc}^{++}$ baryon is observed in the ${\it \Xi}_{c}^{0}  \pi^{+}\pi^{+}$ invariant-mass spectrum.
Using the ${\it \Xi}_{cc}^{++}\to{\it \Lambda}_{c}^{+}K^{-}\pi^{+}\pi^{+}$  decay as the normalisation channel, 
the branching fraction ratio, 
    $\frac{{\cal B}({\it \Xi}_{cc}^{++} \to {\it \Xi}_{c}^{0}\pi^{+}\pi^{+})} {{\cal B}({\it \Xi}_{cc}^{++}\to{\it \Lambda}_{c}^{+} K^{-}\pi^{+}\pi^{+})}$, 
is measured to be \mbox{$1.37\pm 0.18\text{\,(stat)} \pm 0.09\text{\,(syst)} \pm 0.35\text{\,(ext)}$.} This measurement provides critical input for testing QCD factorisation methods in the weak decays of doubly-heavy baryons, particularly in quantifying nonperturbative effects such as final-state interactions and resonance contributions to the hadronisation process.
  
\end{abstract}

\vspace*{2.0cm}

\begin{center}
  Published in JHEP 10 (2025) 136
\end{center}

\vspace{\fill}

{\footnotesize 
\centerline{\copyright~\papercopyright. \href{\paperlicenceurl}{\paperlicence}.}}
\vspace*{2mm}

\end{titlepage}


\newpage
\setcounter{page}{2}
\mbox{~}
%
%
%
%


\renewcommand{\thefootnote}{\arabic{footnote}}
\setcounter{footnote}{0}


\cleardoublepage


\pagestyle{plain} 
\setcounter{page}{1}
\pagenumbering{arabic}


\section{Introduction}
\label{sec:Introduction}

In the quark model~\cite{GellMann:1964nj,Zweig:352337,Zweig:570209}, hadrons are arranged into multiplets according to the symmetry properties of
their wave functions in terms of the flavour, spin and colour quantum numbers of the constituent quarks. 
When four quark flavours, namely \uquark, \dquark, \squark and \cquark, are considered, hadrons form $\mathrm{SU}(4)$ multiplets~\cite{su4_1975}. Within these multiplets, all predicted ground-state baryons with charm quantum numbers $C = 0$ and $C = 1$ have been observed~\cite{PDG2024}, and three weakly decaying baryons with $C = 2$ are expected, $\Xiccpp (ccu)$,  $\Xiccp (ccd)$ and $\Omegacc(ccs)$, collectively referred to as doubly charmed baryons.\footnote{Charge conjugated states are included throughout this paper unless otherwise stated.}
These states provide unique information on the dynamics of low-energy interactions between two heavy quarks in a confined object, yielding new important insights on the theory of quantum chromodynamics (QCD).
 
In 2017, the LHCb collaboration reported the observation of a significant structure in the $\Lc\Km\pip\pip$ mass spectrum, consistent with originating from the doubly charmed baryon $\Xiccpp$ \cite{LHCb-PAPER-2017-018}. 
The $\Xiccpp$ baryon was later confirmed in $\decay{\Xiccpp}{\Xicp\pip}$~\cite{LHCb-PAPER-2018-026} and $\decay{\Xiccpp}{\Xicppr\pip}$ decays~\cite{LHCb-PAPER-2021-052}, while no significant signal was found in a search for $\decay{\Xiccpp}{\Dp\proton\Km\pip}$ decays~\cite{LHCb-PAPER-2019-011}. 
The production cross-section of the $\Xiccpp$ baryon times the branching fraction of the $\decay{\Xiccpp}{\Lc\Km\pip\pip}$ decay has been measured with respect to that of the $\Lc$ baryon, in proton-proton ($pp$) collisions at a centre-of-mass energy of $13 \tev$~\cite{LHCb-PAPER-2019-035}. Its lifetime and mass have been determined to be $0.256^{+0.024}_{-0.022} \stat\pm0.014 \syst \ps$~\cite{LHCb-PAPER-2018-019} and $3621.55\pm0.23 \stat\pm0.30 \syst \mevcc$~\cite{LHCb-PAPER-2019-037}, respectively, which are consistent with a range of theoretical predictions obtained with various methods including heavy quark effective theory (HQET)~\cite{th1}, QCD sum rules~\cite{th3,th6}, perturbative QCD~\cite{th4}, lattice QCD~\cite{th8}, and relativistic quark model~\cite{th10}.
Nevertheless, only three decay modes of the $\Xiccpp$ baryon have been observed so far, although many more have been predicted~\cite{th7,th13,branch_2019,Cheng:2020wmk}. 
Measurements of additional $\Xiccpp$ decay modes provide essential information to better understand the decay dynamics of doubly charmed baryons~\cite{Yu:2019lxw}.

\begin{figure}[!b]
    \includegraphics[width=0.99\textwidth]{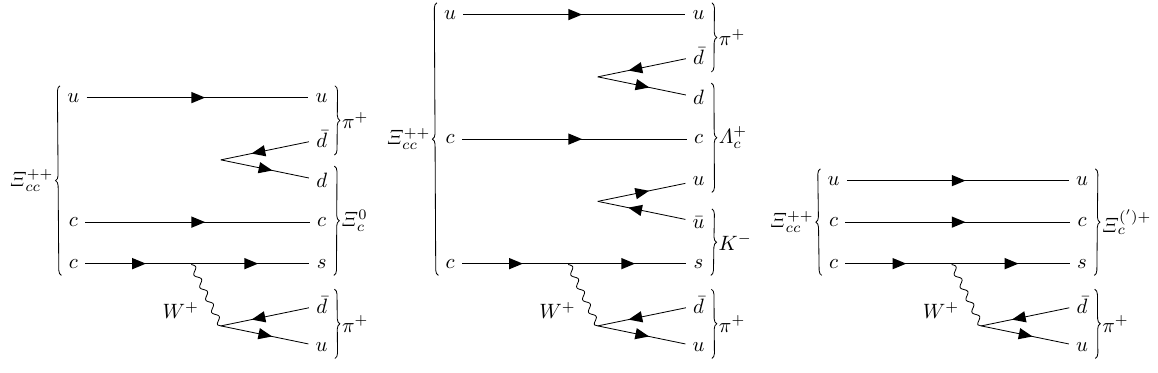}
    \caption{Examples of diagrams contributing to (left) $\mathit{\Xi}_{cc}^{++}\to\mathit{\Xi}_{c}^{0}\pi^{+}\pi^{+}$,
    (middle)  $\mathit{\Xi}_{cc}^{++}\to\mathit{\Lambda}_{c}^{+} K^{-}\pi^{+}\pi^{+}$ and 
    (right) $\mathit{\Xi}_{cc}^{++}\to\mathit{\Xi}_{c}^{(\prime)+}\pi^{+}$ decays.}
    \label{fig:1-1}
\end{figure}

This paper reports the first observation of the $\decay{\Xiccpp}{\Xicz\pip\pip}$ decay in  $\proton\proton$ collision data collected by the~\lhcb experiment at a centre-of-mass energy of $13\tev$, corresponding to an integrated luminosity of $5.4\invfb$. 
The $\decay{\Xiccpp}{\Xicz\pip\pip}$ decay is mediated by the same $\decay{c}{su\Bar{d}}$ weak transition 
as for the known $\decay{\Xiccpp}{\Lc\Km\pip\pip}$ and $\decay{\Xiccpp}{\Xicporpr\pip}$ decays,
as shown in Fig.~\ref{fig:1-1}. However, the formation of additional quark-antiquark pairs and the subsequent hadronisation dynamics are dominated by nonperturbative strong interactions~\cite{Hu_2020mxk,Shi_2019hbf}. Theoretical prediction of the branching fraction is challenged by the interplay of multiple factors: (i) Coupled-channel interference mediated by resonances in the multi-body phase space~\cite{Shi_2017dto}, (ii) limited predictive power of QCD factorisation near the charm-quark mass scale ($m_{\cquark}
\sim 1.5\gevcc$), and (iii) the violation of heavy quark symmetry of HQET in the doubly heavy baryon system~\cite{Shi_2020}. Therefore, measurement of the branching fractions can help to understand the nonperturbative interactions in the weak decays of doubly heavy baryons.

The branching fraction of the $\decay{\Xiccpp}{\Xicz\pip\pip}$ decay is measured relative to the normalisation channel $\decay{\Xiccpp}{\Lc\Km\pip\pip}$. 
The $\Xicz$ and $\Lc$ baryons in the signal and normalisation channels are reconstructed via the $\proton\Km\Km\pip$ and $\proton\Km\pip$ final states, respectively. 
The branching fraction ratio is defined and measured according to
\begin{equation*}
    \mathcal{R}\equiv \frac{\BF(\decay{\Xiccpp}{\Xicz\pip\pip})\times\BF(\decay{\Xicz}{\proton\Km\Km\pip})}
    {\BF(\decay{\Xiccpp}{\Lc\Km\pip\pip})\times\BF(\decay{\Lc}{\proton\Km\pip})}= \frac{N_{\mathrm{sig}}}{N_{\mathrm{norm}}} \times \frac{\varepsilon_{\mathrm{norm}}}{\varepsilon_{\mathrm{sig}}},
    \label{eq:1-ratio}
\end{equation*}
where $\BF$ denotes the branching fraction of the relevant decay mode, and $N_{\mathrm{sig(norm)}}$ and $\varepsilon_\mathrm{sig(norm)}$ are 
the yield and efficiency for the signal (normalisation) channel, respectively. Compared to $\Lc$, the $\Xicz$ baryon has a shorter lifetime, resulting in significantly lower detection efficiency. The smaller branching fraction of $\decay{\Xicz}{\proton\Km\Km\pip}$ makes the study of the $\Xiccpp\to\Xicz\pip\pip$ decay more challenging. However, since the final-state particles in the two $\Xiccpp$ decays are exactly the same, many systematic uncertainties cancel in the ratio of branching fractions. In order to avoid experimenter's bias, the results of the analysis were not examined until the full procedure had been finalised.

\section{Detector, trigger and simulation}
\label{sec:Detector}
The \lhcb detector~\cite{LHCb-DP-2008-001,LHCb-DP-2014-002} is a single-arm forward
spectrometer covering the \mbox{pseudorapidity} range $2<\eta <5$,
designed for the study of particles containing \bquark or \cquark
quarks. The detector includes a high-precision tracking system
consisting of a silicon-strip vertex detector surrounding the $pp$
interaction region~\cite{LHCb-DP-2014-001}, a large-area silicon-strip detector located
upstream of a dipole magnet with a bending power of about
$4{\mathrm{\,T\,m}}$, and three stations of silicon-strip detectors and straw
drift tubes~\cite{LHCb-DP-2017-001}
placed downstream of the magnet.
The tracking system provides a measurement of the momentum, \ptot, of charged particles with
a relative uncertainty that varies from 0.5\% at  $2\gevc$ to 1.0\% at $200\gevc$.
The minimum distance of a track to a primary $pp$ collision vertex (PV), the impact parameter (IP), 
is measured with a resolution of $(15+29/\pt)\mum$,
where \pt is the component of the momentum transverse to the beam, in\,\gevc.
Different types of charged hadrons are distinguished using information
from two ring-imaging Cherenkov (RICH) detectors~\cite{LHCb-DP-2012-003}. 
Photons, electrons and hadrons are identified by a calorimeter system consisting of
scintillating-pad and preshower detectors, an electromagnetic
and a hadronic calorimeter. 
Muons are identified by a system composed of alternating layers of iron and multiwire proportional chambers~\cite{LHCb-DP-2012-002}.

The online event reconstruction and selection are performed by a trigger~\cite{LHCb-DP-2012-004}, 
which consists of a hardware and a software level.
At the hardware level, events are triggered if they contain a muon with high \pt, 
or a hadron, photon, or electron with high transverse energy in the calorimeters.
A $\Xiccpp$ candidate is required to have a positive hardware trigger decision,
either due to particles from the signal decay (Trigger-On-Signal, TOS),
or due to particles in the rest of the event (Trigger-Independent-of-Signal, TIS).
Due to the different size of systematic uncertainties induced by the hardware trigger requirement, candidates are divided into two disjoint categories, denoted as TIS and exTOS, where the latter includes candidates that are selected by the TOS but not by the TIS requirement. The branching fraction ratio is measured independently for each trigger category, and the baseline result comes from the combination of the two measurements.

The software trigger is divided into two stages.
The first stage selects a single particle with large $\pt$ and significant separation from all PVs in the event, or a two-particle vertex displaced from the PV. 
In the second software trigger stage, 
the full event reconstruction and selection is performed~\cite{LHCb-DP-2016-001}.
Using $\proton$, $\Km$ and $\pip$ hadrons that are well identified by the RICH detectors, $\Xicz$ candidates are built by forming $\proton\Km\Km\pip$ particle combinations.
The proton is required to have $p>10\gevc$ to ensure valid particle identification (PID) information~\cite{LHCb-DP-2012-003}. 
All particles in the $\Xicz$ final state must have a good track-fit quality, $\pt>500\mevc$, and $\chisqip>6$ with respect to any PVs, where
$\chisqip$ is defined as the difference in the vertex-fit $\chisq$ of a given PV reconstructed with and without the particle under consideration. 
Additionally, at least one of the four final-state
particles must have $\pt>1\gevc$ and $\chisqip>8$. 
The scalar sum of $\pt$ of the four final-state particles must be greater than $3\gevc$. 
The $\Xicz$ candidate must have a good vertex-fit quality, 
and the vertex must be displaced from the associated PV with a distance corresponding to a decay time of at least $0.1\ps$. The associated PV is the one that best fits the flight direction of the reconstructed candidate. 
Each $\Xicz$ candidate is combined with two $\pip$ mesons to form a $\Xiccpp$ candidate.
The two pions are required to have $\pt>200\mevc$ and good track-fit quality. 
The $\Xiccpp$ candidates must have a good vertex-fit quality, and the vertex of the $\Xicz$ baryon must be located downstream than that of the $\Xiccpp$ baryon.

The selection criteria for the normalisation channel are similar to those for the signal decay. Appropriate  adjustments on the thresholds are made taking into account the differences due to the intermediate $\Lc$ and $\Xicz$ baryons.

The momentum scale is calibrated using samples of $\decay{\jpsi}{\mumu}$ 
and $\decay{\Bu}{\jpsi\Kp}$~decays collected concurrently
with the data sample used for this analysis~\cite{LHCb-PAPER-2012-048,LHCb-PAPER-2013-011}.
The relative accuracy of the calibration is estimated to be $3 \times 10^{-4}$ using samples of other
fully reconstructed decays of beauty hadrons and $\KS$ mesons.

Simulated $\Xiccpp$ decays are used to optimise the event selections, study the signal invariant-mass distributions, and evaluate the efficiencies used for the measurement of the relative branching fraction.
In the simulation, $pp$ collisions are generated using
\pythia~\cite{Sjostrand:2007gs,*Sjostrand:2006za} 
with a specific \lhcb configuration~\cite{LHCb-PROC-2010-056}.
A dedicated generator for doubly heavy baryon production, \genxicc2.0~\cite{genxicc_2010}, is used to produce the $\Xiccpp$ baryons. 
Decays of unstable particles
are described by \evtgen~\cite{Lange:2001uf}, in which final-state radiation is generated using \photos~\cite{davidson2015photos}.
The interaction of the generated particles with the detector, and its response, are implemented using the \geant toolkit~\cite{Allison:2006ve, *Agostinelli:2002hh} as described in Ref.~\cite{LHCb-PROC-2011-006}. 
In the simulation, $\Xiccpp$ baryons are generated with a mass of $3621\mevcc$ and a lifetime of $0.256\ps$~\cite{LHCb-PAPER-2018-019,LHCb-PAPER-2019-037}. The decay products of the $\Xiccpp$ baryons are uniformly distributed in the allowed phase space.
The singly charmed $\Xicz$ decays are distributed according to a pseudoresonant model in which half of the $\Xicz$ decays proceed through the resonant decay $\decay{\Xicz}{\proton\Km\Kstarb(892)^0}$ followed by the decay of the $\Kstarb(892)^0$ meson to the $\Km\pip$ final state~\cite{PDG2024}, while the others are uniformly distributed in the phase space. 
For the normalisation channel, 
the decay products of the $\Xiccpp$ baryons are uniformly distributed in the allowed phase space,
while the $\Lc$ baryons are assumed to decay via a four-component pseudoresonant model,
which includes the phase-space decay, and the $\proton\Kstarb(892)^0$, $\PLambda(1520)\pip$, and $\PDelta(1232)^{++}\Km$ resonant processes.

\section{Event selection}
\label{sec:Selection}
The event selection is performed in three steps: 
a preselection, a multivariate analysis, and the veto of physics background.
In the preselection, 
each particle in the $\Xiccpp$ final state is required to have a momentum within $[2,150]\gevc$ and pseudorapidity within $[1.6,5.3]$.
The $\Xiccpp$ candidate is required to originate from its associated PV. 
The invariant mass of $\proton\Km\Km\pip$ from the $\Xicz$ candidate is required to be within $[2453,2489]\mevcc$, which corresponds to $\pm3$ times the mass resolution around the known mass~\cite{PDG2024}. 
The $\Xiccpp$ candidates reconstructed in the signal and normalisation samples are required to have rapidity within $[2,4.5]$ and transverse momentum within $[4,15]\gevc$.

The combinatorial background is further suppressed with a boosted decision tree~(BDT)~\cite{Breiman,AdaBoost} classifier implemented in the TMVA toolkit~\cite{Hocker:2007ht,*TMVA4}. 
The classifier is trained using simulated decays as a signal proxy and the $\Xicz\pip\pim$ wrong-sign~(WS) sample with the mass in the region of $[3501, 3741]\mevcc$ as a background proxy. 
The BDT training exploits the differences between signal and background candidates in the distributions of the fit qualities of $\Xiccpp$ and $\Xicz$ decay vertices, the $\chisqip$ and $\pt$ of $\Xiccpp$ and $\Xicz$ candidates and their decay products, the angle between the momentum and the vector formed by the PV and the decay vertex for $\Xiccpp$ and $\Xicz$ candidates. The PID information calibrated from data~\cite{LHCb-PUB-2016-021} of all final-state particles is also exploited, which discriminates between pion, kaon and proton particles.
The same procedure is performed for the signal and normalisation decays.

The requirement in the output of the BDT classifier for the signal channel is optimised using the Punzi figure of merit ${\varepsilon_{\rm sig}}/({n_\sigma/2+\sqrt{N_{\rm bkg}}})$~\cite{Punzi:2003bu}, where $\varepsilon_{\rm sig}$ is the signal efficiency obtained from simulation, and $n_\sigma=5$ corresponds to the target significance of five standard deviations. The background yield in the signal region of $\pm 15\mevcc$ around the known $\Xiccpp$ mass~\cite{PDG2024}, denoted as $N_{\rm bkg}$, is estimated by counting the WS yield in the signal region and scaling it by the yield ratio of right-sign and wrong-sign data in the mass region $[3750, 3800]\mevcc$.
The optimal BDT selection corresponds to a signal efficiency of approximately 65\% and a background rejection of approximately 99\%. For the normalisation channel, the working point of the classifier is chosen to have the same BDT efficiency as the signal channel to minimise the systematic uncertainty.

The normalisation decay $\decay{\Xiccpp}{\Lc(}﻿{\to\proton\Km\pip)\Km\pip\pip}$ can be misreconstructed as the signal decay if the \Lc decay products are combined with particles produced promptly from the \Xiccpp decay to form a \Xicz candidate.
This physics background component peaks in the \Xiccpp invariant-mass distribution
and is suppressed to a negligible level by vetoing $\proton\Km\pip$ candidates with invariant masses within $[2268, 2304]\mevcc$, \ie consistent with the known $\Lc$ mass~\cite{PDG2024}.
The efficiency of this requirement for the signal decay is about 80\%, estimated from simulation.
\section{Mass distributions and signal significance}
\label{sec:massfit}
\begin{figure}[!tbp]
    \centering
    \includegraphics[width=0.49\textwidth]{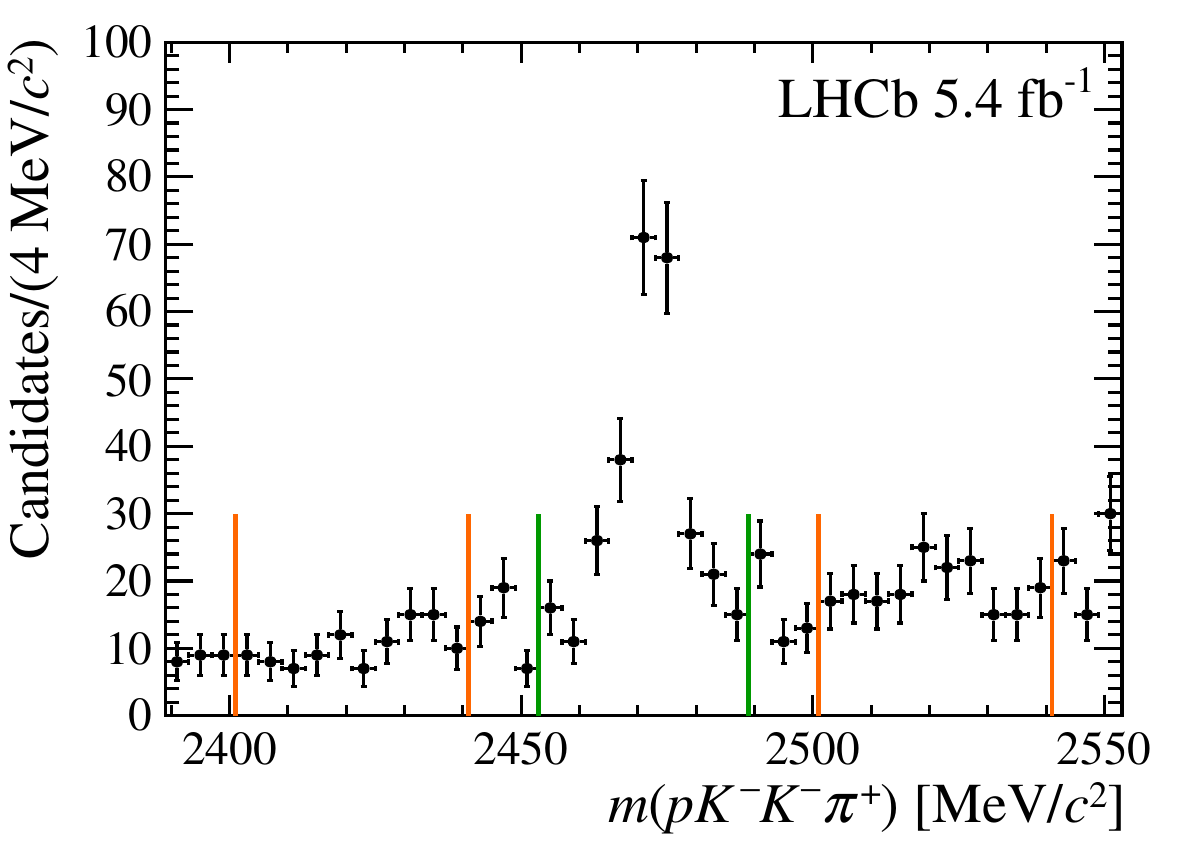}
    \includegraphics[width=0.49\textwidth]{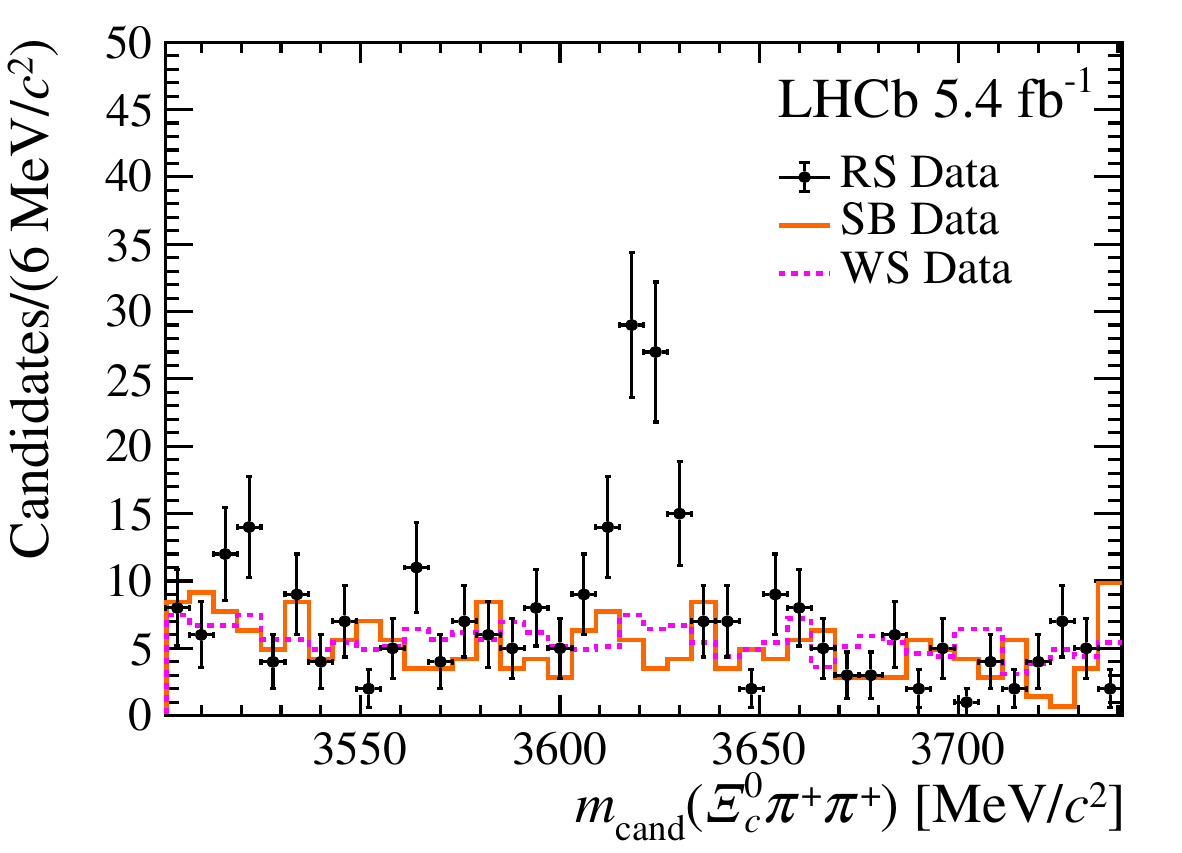}
    \caption{Distributions of the invariant mass of the (left) $p K^{-}K^{-}\pi^{+}$ and (right) $\mathit{\Xi}_{c}^{0}\pi^{+}\pi^{+}$ systems. 
    In the left plot, the boundary of the signal and sideband mass regions of the $\mathit{\Xi}_{c}^{0}$ candidates are indicated by green and orange lines, respectively. In the right plot, the right-sign (RS) signal decays are shown in black points, 
    along with the sample with $\mathit{\Xi}_{c}^{0}$ candidates in the mass sidebands (SB) in pink dashed line and the sample with wrong-sign (WS) combinations in orange solid line, both normalised to have the same area as the RS sample in the $m_{\mathrm{cand}}\left(\mathit{\Xi}_{c}^{0}\pi^{+}\pi^{+}\right)$ sidebands given by $[3501,3561]\cup[3681,3741]\text{\,Me\kern -0.1em V\!/}c^2$.}
    \label{fig:mass}
\end{figure}

In order to improve the mass resolution~\cite{LHCb-PAPER-2019-037}, the invariant mass of the $\Xiccpp$ candidate is measured as
\begin{equation*}
m_{\mathrm{cand}}\left(\Xicz\pip\pip\right)\equiv m\left(\Xicz\pip\pip\right)
    - m\left(\Xicz\right)+M_{\mathrm{PDG}}\left(\Xicz\right),
\end{equation*}
for the signal channel, 
where $m\left(\Xicz\pip\pip\right)$ and $m\left(\Xicz\right)$ are the invariant mass of the $\Xicz\pip\pip$ and  $\proton\Km\Km\pip$ systems, respectively, and $M_{\mathrm{PDG}}\left(\Xicz\right)$ is the known mass of the $\Xicz$ baryon~\cite{PDG2024}.
For the normalisation channel, an analogous definition of 
\begin{equation*}
m_{\mathrm{cand}}\left(\Lc\Km\pip\pip\right)\equiv m\left(\Lc\Km\pip\pip\right)
    - m\left(\Lc\right)+M_{\mathrm{PDG}}\left(\Lc\right),
\end{equation*}
is adopted.
The distributions of the invariant mass of the $\proton\Km\Km\pip$ and the $\Xicz\pip\pip$ systems are shown in Fig.~\ref{fig:mass}.
A significant structure can be seen in the $\Xicz\pip\pip$ invariant-mass distribution at a mass of approximately 3620\mevcc. 
No significant structure is visible in the WS sample, nor in the sample with $\Xicz$ candidates in the mass sidebands given by $[2401,2441]\cup[2501,2541]\mevcc$. 

\begin{figure}[!tbp]
    \centering
    \includegraphics[width=0.49\textwidth]{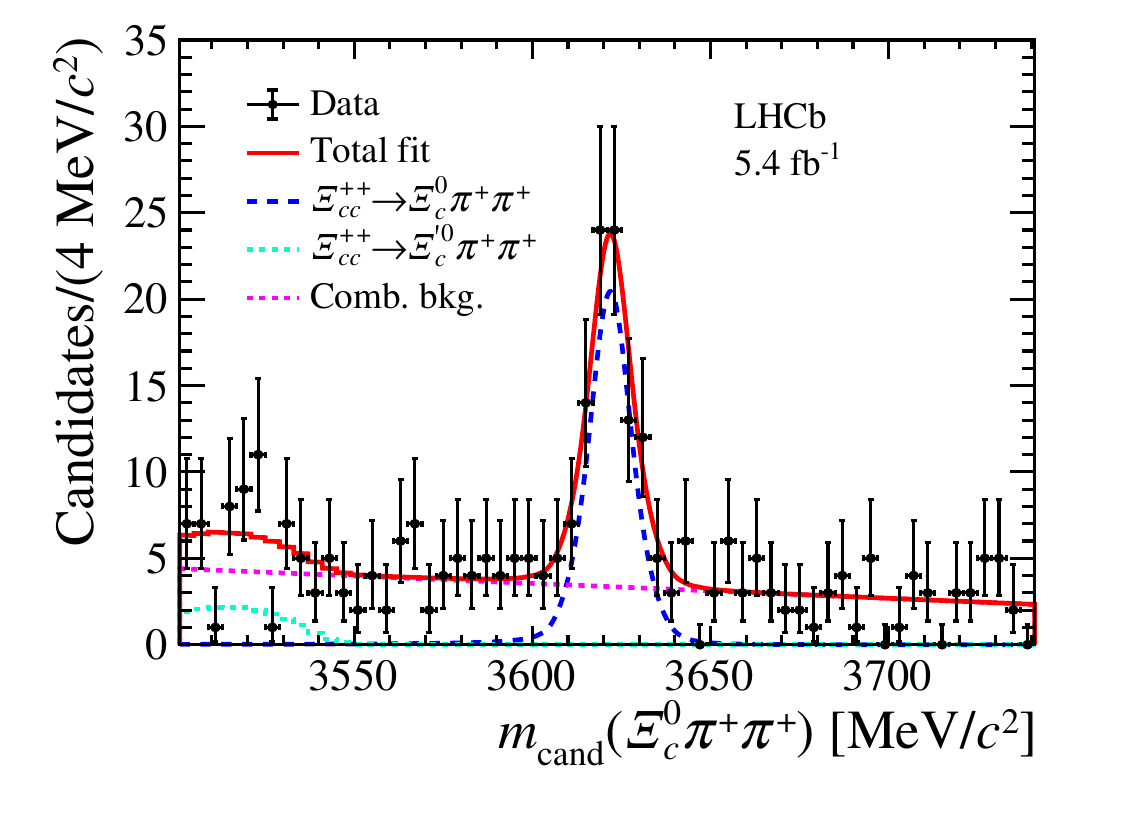}
    \includegraphics[width=0.49\textwidth]{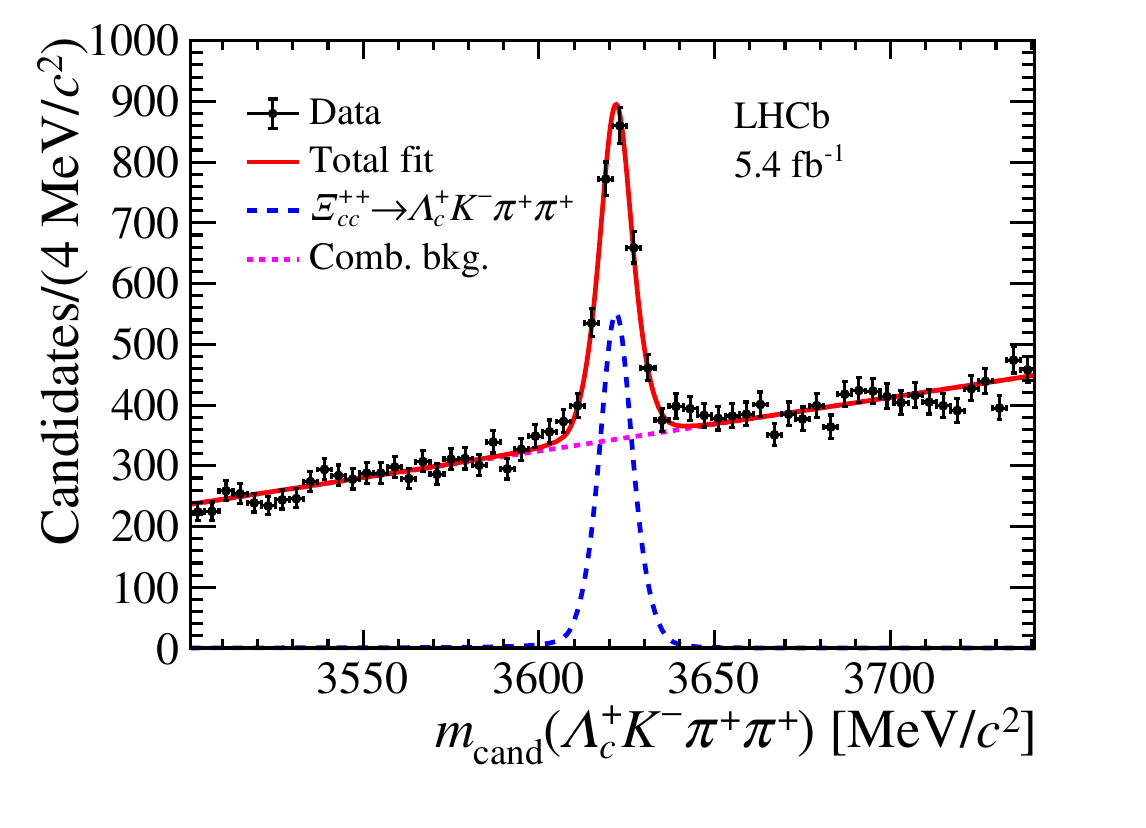}
    \caption{Distributions of the invariant masses of the (left) $\mathit{\Xi}_{c}^{0}\pi^{+}\pi^{+}$ and (right) $\mathit{\Lambda}_{c}^{+}K^{-}\pi^{+}\pi^{+}$ systems, along with the result of the simultaneous fit.}
    \label{fig:fit_part}
\end{figure}

An unbinned maximum-likelihood fit is performed simultaneously to the invariant-mass distributions of the $\Xiccpp$ candidates for the signal and normalisation channels. 
In both channels, the $\Xiccpp$ contribution is described with the sum of a Gaussian function and a double-sided Crystal Ball (DSCB) function~\cite{Skwarnicki:1986xj}, with the fraction of the DSCB and the Gaussian fixed to the value obtained in simulation,
and the combinatorial background is described with a first-order polynomial. 
The shape parameters of the DSCB function are fixed to the values obtained from simulation, except for the peak position and width, while the parameters of the polynomial background function are free to vary.
For the signal channel, there is a contribution from the partially reconstructed $\decay{\Xiccpp}{\PXi_{c}^{\prime 0}(}{\to\Xicz\Pgamma}{)\pip\pip}$ decay, whose shape is obtained using simulated samples generated with RapidSim~\cite{Cowan:2016tnm} and fixed in the fit.
In the simultaneous fit, the signal and normalisation channels share the same peak position, 
and the ratio of the peak widths is fixed to the value obtained from the simulation.
Figure~\ref{fig:fit_part} shows the invariant-mass distributions of the \Xiccpp candidates in both signal and normalisation channels,
along with the fit results.
The significance of the $\decay{\Xiccpp}{\Xicz\pip\pip}$ signal is estimated to be above $10\sigma$ 
according to Wilks' theorem~\cite{significance_2011},
taking into account the systematic uncertainty due to the fit model, while the partially reconstructed $\decay{\Xiccpp}{\PXi_{c}^{\prime 0}(}{\to\Xicz\Pgamma}{)\pip\pip}$ component is not significant.
Simultaneous fits are then performed for the TIS and exTOS trigger categories separately, from which the signal yields are obtained and summarised in Table~\ref{tab:yield}. 

The background-subtracted invariant-mass distributions of the two \Xicz\pip systems in the signal \decay{\Xiccpp}{\Xicz\pip\pip} decay are obtained using the \sPlot~\cite{Pivk:2004ty} method with $m_{\mathrm{cand}}\left(\Xicz\pip\pip\right)$ as the discriminating variable. No significant narrow \Xicp excited states are observed.

\begin{table}[!bp]
    \caption{Yields and efficiencies of the signal and normalisation channels. 
    The uncertainties on the yields are statistical only; the uncertainties on the efficiencies include only the contribution from the limited size of the simulation sample.}
    \centering
    \begin{tabular}{ccccc}
    \toprule
     Category          & $N_{\mathrm{sig}}$ & $N_{\mathrm{norm}}$ & $\varepsilon_{\mathrm{sig}}$ [$\times10^{-4}$] & $\varepsilon_{\mathrm{norm}}$ [$\times10^{-4}$]\\ 
     \midrule
     TIS   & $62\pm9$  & $1279\pm55$ &$1.159\pm0.023$ & $2.547\pm0.090$\\          
     exTOS & $21\pm6$  & $\phantom0 461\pm34$ &$0.286\pm0.011$ & $0.624\pm0.034$ \\ 
     \bottomrule
    \end{tabular}
    \label{tab:yield}
\end{table}

\section{Efficiencies}
\label{sec:Efficiency}
Simulated samples of the signal and normalisation channels are used to evaluate the reconstruction and selection efficiencies.
Corrections are applied to these samples to ensure that the quantities which affect the efficiency determination are well described by the simulation.
Namely, the charged particle multiplicity and the $\Xiccpp$ transverse-momentum distributions of the simulation sample are corrected to match those of the normalisation channel data.
In addition, track-momentum- and pseudorapidity-dependent corrections are applied to better match the PID response and tracking efficiency in data, as described in  Refs.~\cite{LHCb-PUB-2016-021, LHCb-DP-2013-002}. 
The phase-space distributions of the $\decay{\Xiccpp}{\Lc\Km\pip\pip}$, $\decay{\Lc}{\proton\Km\pip}$ and $\decay{\Xicz}{\proton\Km\Km\pip}$ decays in the simulation samples are weighted to match the data~\cite{Rogozhnikov:2016bdp}. The $\decay{\Xiccpp}{\Lc\Km\pip\pip}$ and $\decay{\Lc}{\proton\Km\pip}$ phase-space distributions in the data are taken from the normalisation decay, while the $\decay{\Xicz}{\proton\Km\Km\pip}$ phase-space distributions is taken from that of the $\Xicz$ baryon in the abundant $\decay{\Xibm}{\Xicz\pim}$ decay~\cite{LHCb-PAPER-2016-008}.
After implementing these corrections, the simulation and data for the signal channel show a high degree of agreement. However, a discrepancy remains in the efficiency of the BDT selection for the normalisation channel. To investigate this, a looser BDT selection is applied while maintaining a discernible peak in the $m_{\mathrm{cand}}\left(\Lc\Km\pip\pip\right)$ spectrum of the data. This discrepancy is corrected by multiplying the efficiency for the normalisation channel by a factor,
\begin{equation*}
    f_{\rm BDT}=\frac{\varepsilon_{\rm data}(\rm Nominal\ BDT|Looser\ BDT)}{\varepsilon_{\rm sim}(\rm Nominal\ BDT|Looser\ BDT)},
\end{equation*}
where $\varepsilon_{\rm data(sim)}(\rm Nominal\ BDT|Looser\ BDT)$ is the efficiency of the nominal BDT requirement, given that the looser BDT requirement has been applied, for the data (simulation) samples.
The correction factor is determined to be $0.90\pm0.04$, where the uncertainty is systematic as discussed in Sec.~\ref{sec:Systematic}.
The background-subtracted distributions in data are determined with the \sPlot method.

The total efficiencies after corrections for the signal and normalisation channels 
and for the two hardware trigger categories are shown in Table~\ref{tab:yield},
from which the efficiency ratios between the signal and the normalisation channel are
determined to be $0.454\pm0.019$ and $0.458\pm0.030$ for the TIS and exTOS categories, respectively.
The uncertainties are attributed to the limited size of the simulation samples only.

\section{Systematic uncertainties}
\label{sec:Systematic}
The branching fraction ratio measurement has systematic uncertainties arising from determinations of the relative yields and relative efficiencies, as summarised in Table~\ref{tab:syst} and described below.
Individual systematic uncertainties are added in quadrature to obtain the total uncertainty, and determined to be 7.0\% and 12.5\% for the TIS and exTOS samples, respectively. 
\begin{table}[!b]
    \centering
    \caption{Summary of the relative systematic uncertainties on the branching fraction ratio $\mathcal{R}$.}
    \label{tab:syst}
    \begin{tabular}{lcc}
    \toprule
    Source    &  TIS(\%) & exTOS(\%)\\
    \midrule
    Fit model       &1.1   &1.1 \\
    Simulation sample size   &4.2   &6.6 \\
    \Xiccpp lifetime &1.0   &1.0 \\
    Kinematic correction        &1.2    &1.2   \\ 
    Decay dynamics    &2.0  &2.0  \\
    Tracking         &1.5   &1.5 \\
    Particle identification  &1.4   &1.4 \\
    Hardware trigger       &---        &9.0\\
    Hit error parametrisation  &0.5    &0.5   \\  
    BDT efficiency correction   &4.4   &4.4\\
    
    \midrule
    Total           &7.0   &12.5 \\
    \bottomrule
    \end{tabular}
\end{table}

The models used in the fits to the invariant-mass distributions may bias
the relative signal yields. 
This systematic uncertainty is studied by using alternative functions to describe the shapes of the $\Xiccpp$ and background components simultaneously for the signal and normalisation channels. 
The sum of two Gaussian functions is chosen as the $\Xiccpp$ model, the second-order polynomial function is chosen as the alternative combinatorial background model, and a third alternative model is the baseline model excluding the partially reconstructed component of the signal channel. The change in the yield ratio is assigned as the systematic uncertainty for each alternative model, and the sum in quadrature of the three sources is 1.1\%.

Three categories of systematic uncertainty arising from the determination of the relative efficiencies are evaluated. 
First, the uncertainty due to the limited size of the simulated samples is evaluated to be 4.2\% and 6.6\% for the TIS and the exTOS result, respectively. 

The second category is due to the imperfect or missing knowledge of the $\Xiccpp$ baryon properties, including its lifetime and  kinematic distributions, and the resonant structure of  $\decay{\Xiccpp}{\Xicz\pip\pip}$ and $\decay{\Xiccpp}{\Lc\Km\pip\pip}$ decays.
The uncertainty in the selection efficiency due to the imprecise $\Xiccpp$ lifetime~\cite{LHCb-PAPER-2018-019,PDG2024} is evaluated by weighting the simulation samples to match the known $\Xiccpp$ lifetime values varied within $\pm 1$ standard deviation.
The largest relative variation in the efficiency ratio, 1.0\%, is assigned as a systematic uncertainty. 
The systematic uncertainty associated with the $\Xiccpp$ kinematic corrections is tested with simulations. For each test, the simulated samples are weighted to match a specific kinematic distribution, where the correction factor in each transverse-momentum interval of the $\Xiccpp$ baryon and the charged particle multiplicity interval is varied according to a Gaussian distribution defined by the nominal value and its uncertainty. 
The width of the resulting distribution of the efficiency ratio in different tests is assigned as the corresponding systematic uncertainty, which is 1.2\%.
The dynamics of the $\decay{\Xiccpp}{\Xicz\pip\pip}$ decay is characterised by the resonant structures and the polarisation of each quasi-two-body decay chain.
Variation of the efficiency due to possible resonant enhancements is studied by measuring the efficiency in regions of Dalitz-plot variables $\left(m_l(\Xicz\pi^+), m_h(\Xicz\pi^+)\right)$ and performing event-by-event efficiency corrections using the two-dimensional efficiency map, where the subscript $l$ ($h$) refers to the lower (higher) value of the two $\Xicz\pip$ invariant masses.
The relative difference between the nominal efficiency and the per-event-corrected efficiency is taken as the uncertainty due to resonant structures.
The systematic uncertainty associated with the polarisation is also studied with simulations. 
The $\decay{\Xiccpp}{\Xicz(\to\proton\Km\Km\pip)\pip\pip}$ decay can be expressed as the coherent sum of sequential two-body decays~\cite{Lindenbaum:1957ec}. A series of tests are performed using the fully simulated samples weighted to match various polarisations, in which the helicity amplitudes of each two-body decay chain are randomly assigned. The width of the resulting distribution of the efficiencies is taken as the systematic uncertainty due to the unknown polarisation.
The systematic uncertainty due to the decay dynamics is the sum in quadrature of the uncertainties from the resonant structures and the polarisation, and amounts to 2.0\%.

The third category of systematic uncertainty accounts for the discrepancy between data and simulation, including effects due to tracking, particle identification, the hardware trigger, the hit error parametrisation of the vertex detector, and the BDT efficiency. 
The systematic uncertainty associated with the tracking efficiency stems from the limited size of the calibration data samples used to measure it~\cite{LHCb-DP-2013-002}, and is evaluated to be 1.5\%.
The systematic uncertainty associated with the PID correction is evaluated using alternative calibration samples in the PID correction procedure~\cite{LHCb-PUB-2016-021}. The maximum relative change of the branching fraction ratio, 1.4\%, is assigned as the uncertainty from this procedure.
The inconsistency of hit error parametrization of the vertex detector between simulation and data leads to a small discrepancy in $\chisqip$ distributions, and hence biases the efficiency. 
The impact on the efficiency ratio is assessed by scaling the $\chisqip$ values of the final-state particles in simulation using a factor estimated conservatively based on the observed behaviour of the vertex detector. 
The relative change in the efficiency ratio between the signal and normalisation channels is 0.5\%, which is taken as the systematic uncertainty.
The uncertainty due to the simulation of the hardware trigger is evaluated for TIS and exTOS data samples, separately. 
For the exTOS sample, a data-driven method is used to determine the TOS efficiency as a function of the energy deposited in the hadronic calorimeter by each final-state particle~\cite{LHCb-DP-2012-004,
LHCb-DP-2020-001}. 
The difference in the ratio of TOS efficiencies between the baseline value and the one obtained with the data-driven method, 9.0\%, is taken as the systematic uncertainty for the exTOS sample.
The TIS sample is independent of the $\Xiccpp$ decay by construction. Since the kinematic of $\Xiccpp$ and the charged particle multiplicity have been corrected and corresponding systematic uncertainties are studied, no extra systematic uncertainty is assigned to this sample.
In the baseline procedure, a looser BDT requirement is used to correct the BDT efficiency of the normalisation channel. However, the efficiency of this looser BDT requirement may differ between the simulation and the data. To investigate the uncertainty associated with this procedure, a traditional selection method without machine learning is used in place of the looser BDT requirement to derive the correction factor. The difference between the baseline value and the one obtained with the alternative selection, 4.4\%, is taken as the uncertainty due to the correction of the BDT efficiency.

\section{Result and conclusion}
\label{sec:Conclusion}
Taking into account the ratios of the $\Xiccpp$ yields and efficiencies between the signal and normalisation channels, 
and associated uncertainties, 
the branching fraction ratio is measured to be 
\begin{equation*}
\begin{aligned}
\mathcal{R} &= 0.107 \pm 0.016 \text{(stat)} \pm 0.007 \text{(syst)},\text{ for the TIS sample},\\
    \mathcal{R} &= 0.100 \pm 0.029 \text{(stat)} \pm 0.012 \text{(syst)},\text{ for the exTOS sample}.
    \end{aligned}
\end{equation*}

The measurements from the two samples are consistent and are combined using the best linear unbiased estimator~\cite{Nisius:2014wua,Nisius:2020jmf} to obtain
\begin{equation*}
\mathcal{R} 
    = 0.105 \pm 0.014 \text{(stat)} \pm 0.007 \text{(syst)}.
\end{equation*}
In the combination, systematic uncertainties arising from the limited size of the simulated samples are assumed to be uncorrelated between the TIS and exTOS categories,
while the remaining nonzero systematic uncertainties are taken to be fully correlated.

Using the measured branching fractions of $\decay{\Lc}{\proton\Km\pip}$ and $\decay{\Xicz}{\proton\Km\Km\pip}$ decays~\cite{PDG2024}, the branching fraction ratio of the two doubly charmed baryon decays is measured to be
\begin{equation*}
    \frac{\BF(\decay{\Xiccpp}{\Xicz\pip\pip})}
    {\BF(\decay{\Xiccpp}{\Lc\Km\pip\pip})}= 1.37 \pm 0.18 (\mathrm{stat}) \pm 0.09(\mathrm{syst})\pm 0.35(\mathrm{ext}),
    \label{eq:10-ratio}
\end{equation*}
where the external uncertainty comes from those on the branching fractions of the \mbox{$\decay{\Lc}{\proton\Km\pip}$} and $\decay{\Xicz}{\proton\Km\Km\pip}$ decays~\cite{PDG2024}. Taking into account the contributions of two, three and four-body decays~\cite{Cheng:2020wmk,branch_2019,Wang_2017mqp,Shi_2017dto,Li_2021rfj}, the measured relative branching fraction is consistent with theoretical predictions. Moreover, unlike the $\decay{\Xiccpp}{\Lc\Km\pip\pip}$ decay, the $\decay{\Xiccpp}{\Xicz\pip\pip}$ decay is found to be almost a nonresonant process, and thus is a good channel to test the QCD factorisation method in the three-body doubly-heavy-baryon decay.

In summary, a new mode of the doubly-charmed-baryon decay, $\decay{\Xiccpp}{\Xicz\pip\pip}$, is observed in a data sample of $pp$ collisions collected by the LHCb experiment at a centre-of-mass energy of $13\tev$, corresponding to an integrated luminosity of $5.4\invfb$. 
The branching fraction ratio between the $\decay{\Xiccpp}{\Xicz\pip\pip}$ and $\decay{\Xiccpp}{\Lc\Km\pip\pip}$ decays is measured for the first time. This measurement provides important information towards an improved understanding of the decays of doubly charmed baryons, and contributes to shed light on the dynamics of low-energy interactions between two confined heavy quarks.

\section*{Acknowledgements}
%
%
\noindent We express our gratitude to our colleagues in the CERN
accelerator departments for the excellent performance of the LHC. We
thank the technical and administrative staff at the LHCb
institutes.
We acknowledge support from CERN and from the national agencies:
ARC (Australia);
CAPES, CNPq, FAPERJ and FINEP (Brazil); 
MOST and NSFC (China); 
CNRS/IN2P3 (France); 
BMBF, DFG and MPG (Germany); 
INFN (Italy); 
NWO (Netherlands); 
MNiSW and NCN (Poland); 
MCID/IFA (Romania); 
MICIU and AEI (Spain);
SNSF and SER (Switzerland); 
NASU (Ukraine); 
STFC (United Kingdom); 
DOE NP and NSF (USA).
We acknowledge the computing resources that are provided by ARDC (Australia), 
CBPF (Brazil),
CERN, 
IHEP and LZU (China),
IN2P3 (France), 
KIT and DESY (Germany), 
INFN (Italy), 
SURF (Netherlands),
Polish WLCG (Poland),
IFIN-HH (Romania), 
PIC (Spain), CSCS (Switzerland), 
and GridPP (United Kingdom).
We are indebted to the communities behind the multiple open-source
software packages on which we depend.
Individual groups or members have received support from
Key Research Program of Frontier Sciences of CAS, CAS PIFI, CAS CCEPP, 
Fundamental Research Funds for the Central Universities,  and Sci.\ \& Tech.\ Program of Guangzhou (China);
Minciencias (Colombia);
EPLANET, Marie Sk\l{}odowska-Curie Actions, ERC and NextGenerationEU (European Union);
A*MIDEX, ANR, IPhU and Labex P2IO, and R\'{e}gion Auvergne-Rh\^{o}ne-Alpes (France);
Alexander-von-Humboldt Foundation (Germany);
ICSC (Italy); 
Severo Ochoa and Mar\'ia de Maeztu Units of Excellence, GVA, XuntaGal, GENCAT, InTalent-Inditex and Prog.~Atracci\'on Talento CM (Spain);
SRC (Sweden);
the Leverhulme Trust, the Royal Society and UKRI (United Kingdom).

\addcontentsline{toc}{section}{References}
\bibliographystyle{LHCb}
\bibliography{main,standard,LHCb-PAPER,LHCb-CONF,LHCb-DP,LHCb-TDR}

\ifx\mcitethebibliography\mciteundefinedmacro
\PackageError{LHCb.bst}{mciteplus.sty has not been loaded}
{This bibstyle requires the use of the mciteplus package.}\fi
\providecommand{\href}[2]{#2}
\begin{mcitethebibliography}{10}
\mciteSetBstSublistMode{n}
\mciteSetBstMaxWidthForm{subitem}{\alph{mcitesubitemcount})}
\mciteSetBstSublistLabelBeginEnd{\mcitemaxwidthsubitemform\space}
{\relax}{\relax}

\bibitem{GellMann:1964nj}
M.~Gell-Mann, \ifthenelse{\boolean{articletitles}}{\emph{{A schematic model of baryons and mesons}}, }{}\href{https://doi.org/10.1016/S0031-9163(64)92001-3}{Phys.\ Lett.\  \textbf{8} (1964) 214}\relax
\mciteBstWouldAddEndPuncttrue
\mciteSetBstMidEndSepPunct{\mcitedefaultmidpunct}
{\mcitedefaultendpunct}{\mcitedefaultseppunct}\relax
\EndOfBibitem
\bibitem{Zweig:352337}
G.~Zweig, \ifthenelse{\boolean{articletitles}}{\emph{{An SU$_3$ model for strong interaction symmetry and its breaking; Version 1}}}{}  \href{http://cds.cern.ch/record/352337}{CERN-TH-401}, CERN, Geneva, 1964\relax
\mciteBstWouldAddEndPuncttrue
\mciteSetBstMidEndSepPunct{\mcitedefaultmidpunct}
{\mcitedefaultendpunct}{\mcitedefaultseppunct}\relax
\EndOfBibitem
\bibitem{Zweig:570209}
G.~Zweig, \ifthenelse{\boolean{articletitles}}{\emph{{An SU$_3$ model for strong interaction symmetry and its breaking; Version 2}}}{}  \href{http://cds.cern.ch/record/570209}{CERN-TH-412}, CERN, Geneva, 1964\relax
\mciteBstWouldAddEndPuncttrue
\mciteSetBstMidEndSepPunct{\mcitedefaultmidpunct}
{\mcitedefaultendpunct}{\mcitedefaultseppunct}\relax
\EndOfBibitem
\bibitem{su4_1975}
A.~De~R\'ujula, H.~Georgi, and S.~L. Glashow, \ifthenelse{\boolean{articletitles}}{\emph{Hadron masses in a gauge theory}, }{}\href{https://doi.org/10.1103/PhysRevD.12.147}{Phys.\ Rev.\  \textbf{D12} (1975) 147}\relax
\mciteBstWouldAddEndPuncttrue
\mciteSetBstMidEndSepPunct{\mcitedefaultmidpunct}
{\mcitedefaultendpunct}{\mcitedefaultseppunct}\relax
\EndOfBibitem
\bibitem{PDG2024}
Particle Data Group, S.~Navas {\em et~al.}, \ifthenelse{\boolean{articletitles}}{\emph{{\href{http://pdg.lbl.gov/}{Review of particle physics}}}, }{}\href{https://doi.org/10.1103/PhysRevD.110.030001}{Phys.\ Rev.\  \textbf{D110} (2024) 030001}\relax
\mciteBstWouldAddEndPuncttrue
\mciteSetBstMidEndSepPunct{\mcitedefaultmidpunct}
{\mcitedefaultendpunct}{\mcitedefaultseppunct}\relax
\EndOfBibitem
\bibitem{LHCb-PAPER-2017-018}
LHCb collaboration, R.~Aaij {\em et~al.}, \ifthenelse{\boolean{articletitles}}{\emph{{Observation of the doubly charmed baryon \Xiccpp}}, }{}\href{https://doi.org/10.1103/PhysRevLett.119.112001}{Phys.\ Rev.\ Lett.\  \textbf{119} (2017) 112001}, \href{http://arxiv.org/abs/1707.01621}{{\normalfont\ttfamily arXiv:1707.01621}}\relax
\mciteBstWouldAddEndPuncttrue
\mciteSetBstMidEndSepPunct{\mcitedefaultmidpunct}
{\mcitedefaultendpunct}{\mcitedefaultseppunct}\relax
\EndOfBibitem
\bibitem{LHCb-PAPER-2018-026}
LHCb collaboration, R.~Aaij {\em et~al.}, \ifthenelse{\boolean{articletitles}}{\emph{{First observation of the doubly charmed baryon decay \mbox{\decay{\Xiccpp}{\Xires_c^+\pip}}}}, }{}\href{https://doi.org/10.1103/PhysRevLett.121.162002}{Phys.\ Rev.\ Lett.\  \textbf{121} (2018) 162002}, \href{http://arxiv.org/abs/1807.01919}{{\normalfont\ttfamily arXiv:1807.01919}}\relax
\mciteBstWouldAddEndPuncttrue
\mciteSetBstMidEndSepPunct{\mcitedefaultmidpunct}
{\mcitedefaultendpunct}{\mcitedefaultseppunct}\relax
\EndOfBibitem
\bibitem{LHCb-PAPER-2021-052}
LHCb collaboration, R.~Aaij {\em et~al.}, \ifthenelse{\boolean{articletitles}}{\emph{{Observation of the doubly charmed baryon decay $\Xires_{cc}^{++} \to \Xires_{c}^{\prime +} \pi^{+}$}}, }{}\href{https://doi.org/10.1007/JHEP05(2022)038}{JHEP \textbf{05} (2022) 038}, \href{http://arxiv.org/abs/2202.05648}{{\normalfont\ttfamily arXiv:2202.05648}}\relax
\mciteBstWouldAddEndPuncttrue
\mciteSetBstMidEndSepPunct{\mcitedefaultmidpunct}
{\mcitedefaultendpunct}{\mcitedefaultseppunct}\relax
\EndOfBibitem
\bibitem{LHCb-PAPER-2019-011}
LHCb collaboration, R.~Aaij {\em et~al.}, \ifthenelse{\boolean{articletitles}}{\emph{{A search for \mbox{\decay{\Xiccpp}{\Dp p\Km\pip}} decays}}, }{}\href{https://doi.org/10.1007/JHEP10(2019)124}{JHEP \textbf{10} (2019) 124}, \href{http://arxiv.org/abs/1905.02421}{{\normalfont\ttfamily arXiv:1905.02421}}\relax
\mciteBstWouldAddEndPuncttrue
\mciteSetBstMidEndSepPunct{\mcitedefaultmidpunct}
{\mcitedefaultendpunct}{\mcitedefaultseppunct}\relax
\EndOfBibitem
\bibitem{LHCb-PAPER-2019-035}
LHCb collaboration, R.~Aaij {\em et~al.}, \ifthenelse{\boolean{articletitles}}{\emph{{Measurement of \Xiccpp production in $pp$ collisions at $\sqrt{s} = 13$~\tev}}, }{}\href{https://doi.org/10.1088/1674-1137/44/2/022001}{Chin.\ Phys.\  \textbf{C44} (2020) 022001}, \href{http://arxiv.org/abs/1910.11316}{{\normalfont\ttfamily arXiv:1910.11316}}\relax
\mciteBstWouldAddEndPuncttrue
\mciteSetBstMidEndSepPunct{\mcitedefaultmidpunct}
{\mcitedefaultendpunct}{\mcitedefaultseppunct}\relax
\EndOfBibitem
\bibitem{LHCb-PAPER-2018-019}
LHCb collaboration, R.~Aaij {\em et~al.}, \ifthenelse{\boolean{articletitles}}{\emph{{Measurement of the lifetime of the doubly charmed baryon \Xiccpp}}, }{}\href{https://doi.org/10.1103/PhysRevLett.121.052002}{Phys.\ Rev.\ Lett.\  \textbf{121} (2018) 052002}, \href{http://arxiv.org/abs/1806.02744}{{\normalfont\ttfamily arXiv:1806.02744}}\relax
\mciteBstWouldAddEndPuncttrue
\mciteSetBstMidEndSepPunct{\mcitedefaultmidpunct}
{\mcitedefaultendpunct}{\mcitedefaultseppunct}\relax
\EndOfBibitem
\bibitem{LHCb-PAPER-2019-037}
LHCb collaboration, R.~Aaij {\em et~al.}, \ifthenelse{\boolean{articletitles}}{\emph{{Precision measurement of the $\Xiccpp$ mass}}, }{}\href{https://doi.org/10.1007/JHEP02(2020)049}{JHEP \textbf{02} (2020) 049}, \href{http://arxiv.org/abs/1911.08594}{{\normalfont\ttfamily arXiv:1911.08594}}\relax
\mciteBstWouldAddEndPuncttrue
\mciteSetBstMidEndSepPunct{\mcitedefaultmidpunct}
{\mcitedefaultendpunct}{\mcitedefaultseppunct}\relax
\EndOfBibitem
\bibitem{th1}
W.~Robert and M.~Pervin, \ifthenelse{\boolean{articletitles}}{\emph{Heavy baryons in a quark model}, }{}\href{https://doi.org/10.1142/S0217751X08041219}{Int.\ J.\ Mod.\ Phys.\  \textbf{A23} (2008) 2817}, \href{http://arxiv.org/abs/0711.2492}{{\normalfont\ttfamily arXiv:0711.2492}}\relax
\mciteBstWouldAddEndPuncttrue
\mciteSetBstMidEndSepPunct{\mcitedefaultmidpunct}
{\mcitedefaultendpunct}{\mcitedefaultseppunct}\relax
\EndOfBibitem
\bibitem{th3}
Z.-G. Wang, \ifthenelse{\boolean{articletitles}}{\emph{Analysis of the {$\frac{1}{2}^+$} doubly heavy baryon states with {QCD} sum rules}, }{}\href{https://doi.org/10.1140/epja/i2010-11004-3}{Eur.\ Phys.\ J.\  \textbf{A45} (2010) 267}, \href{http://arxiv.org/abs/1001.4693}{{\normalfont\ttfamily arXiv:1001.4693}}\relax
\mciteBstWouldAddEndPuncttrue
\mciteSetBstMidEndSepPunct{\mcitedefaultmidpunct}
{\mcitedefaultendpunct}{\mcitedefaultseppunct}\relax
\EndOfBibitem
\bibitem{th6}
J.-R. Zhang and M.-Q. Huang, \ifthenelse{\boolean{articletitles}}{\emph{Doubly heavy baryons in {QCD} sum rules}, }{}\href{https://doi.org/10.1103/PhysRevD.78.094007}{Phys.\ Rev.\  \textbf{D78} (2008) 094007}, \href{http://arxiv.org/abs/0810.5396}{{\normalfont\ttfamily arXiv:0810.5396}}\relax
\mciteBstWouldAddEndPuncttrue
\mciteSetBstMidEndSepPunct{\mcitedefaultmidpunct}
{\mcitedefaultendpunct}{\mcitedefaultseppunct}\relax
\EndOfBibitem
\bibitem{th4}
C.-H. Chang, C.-F. Qiao, J.-X. Wang, and X.-G. Wu, \ifthenelse{\boolean{articletitles}}{\emph{Estimate of the hadronic production of the doubly charmed baryon {\Xicc} in the general-mass variable-flavor-number scheme}, }{}\href{https://doi.org/10.1103/PhysRevD.73.094022}{Phys.\ Rev.\  \textbf{D73} (2006) 094022}, \href{http://arxiv.org/abs/hep-ph/0601032}{{\normalfont\ttfamily arXiv:hep-ph/0601032}}\relax
\mciteBstWouldAddEndPuncttrue
\mciteSetBstMidEndSepPunct{\mcitedefaultmidpunct}
{\mcitedefaultendpunct}{\mcitedefaultseppunct}\relax
\EndOfBibitem
\bibitem{th8}
C.~Alexandrou and C.~Kallidonis, \ifthenelse{\boolean{articletitles}}{\emph{Low-lying baryon masses using ${N}_{f}=2$ twisted mass clover-improved fermions directly at the physical pion mass}, }{}\href{https://doi.org/10.1103/PhysRevD.96.034511}{Phys.\ Rev.\  \textbf{D96} (2017) 034511}, \href{http://arxiv.org/abs/1704.02647}{{\normalfont\ttfamily arXiv:1704.02647}}\relax
\mciteBstWouldAddEndPuncttrue
\mciteSetBstMidEndSepPunct{\mcitedefaultmidpunct}
{\mcitedefaultendpunct}{\mcitedefaultseppunct}\relax
\EndOfBibitem
\bibitem{th10}
D.~Ebert, R.~N. Faustov, V.~O. Galkin, and A.~P. Martynenko, \ifthenelse{\boolean{articletitles}}{\emph{Mass spectra of doubly heavy baryons in the relativistic quark model}, }{}\href{https://doi.org/10.1103/PhysRevD.66.014008}{Phys.\ Rev.\  \textbf{D66} (2002) 014008}, \href{http://arxiv.org/abs/hep-ph/0201217}{{\normalfont\ttfamily arXiv:hep-ph/0201217}}\relax
\mciteBstWouldAddEndPuncttrue
\mciteSetBstMidEndSepPunct{\mcitedefaultmidpunct}
{\mcitedefaultendpunct}{\mcitedefaultseppunct}\relax
\EndOfBibitem
\bibitem{th7}
M.~Karliner and J.~L. Rosner, \ifthenelse{\boolean{articletitles}}{\emph{Baryons with two heavy quarks: masses, production, decays, and detection}, }{}\href{https://doi.org/10.1103/PhysRevD.90.094007}{Phys.\ Rev.\  \textbf{D90} (2014) 094007}, \href{http://arxiv.org/abs/1408.5877}{{\normalfont\ttfamily arXiv:1408.5877}}\relax
\mciteBstWouldAddEndPuncttrue
\mciteSetBstMidEndSepPunct{\mcitedefaultmidpunct}
{\mcitedefaultendpunct}{\mcitedefaultseppunct}\relax
\EndOfBibitem
\bibitem{th13}
V.~V. Kiselev and A.~K. Likhoded, \ifthenelse{\boolean{articletitles}}{\emph{{Baryons with two heavy quarks}}, }{}\href{https://doi.org/10.1070/PU2002v045n05ABEH000958}{Phys.\ Usp.\  \textbf{45} (2002) 455}, \href{http://arxiv.org/abs/hep-ph/0103169}{{\normalfont\ttfamily arXiv:hep-ph/0103169}}\relax
\mciteBstWouldAddEndPuncttrue
\mciteSetBstMidEndSepPunct{\mcitedefaultmidpunct}
{\mcitedefaultendpunct}{\mcitedefaultseppunct}\relax
\EndOfBibitem
\bibitem{branch_2019}
T.~Gutsche {\em et~al.}, \ifthenelse{\boolean{articletitles}}{\emph{{Analysis of the semileptonic and nonleptonic two-body decays of the double heavy charm baryon states {\Xiccpp, \Xiccp and \Omegacc}}}, }{}\href{https://doi.org/10.1103/PhysRevD.100.114037}{Phys.\ Rev.\  \textbf{D100} (2019) 114037}, \href{http://arxiv.org/abs/1911.10785}{{\normalfont\ttfamily arXiv:1911.10785}}\relax
\mciteBstWouldAddEndPuncttrue
\mciteSetBstMidEndSepPunct{\mcitedefaultmidpunct}
{\mcitedefaultendpunct}{\mcitedefaultseppunct}\relax
\EndOfBibitem
\bibitem{Cheng:2020wmk}
H.-Y. Cheng, G.~Meng, F.~Xu, and J.~Zou, \ifthenelse{\boolean{articletitles}}{\emph{{Two-body weak decays of doubly charmed baryons}}, }{}\href{https://doi.org/10.1103/PhysRevD.101.034034}{Phys.\ Rev.\  \textbf{D101} (2020) 034034}, \href{http://arxiv.org/abs/2001.04553}{{\normalfont\ttfamily arXiv:2001.04553}}\relax
\mciteBstWouldAddEndPuncttrue
\mciteSetBstMidEndSepPunct{\mcitedefaultmidpunct}
{\mcitedefaultendpunct}{\mcitedefaultseppunct}\relax
\EndOfBibitem
\bibitem{Yu:2019lxw}
F.-S. Yu, \ifthenelse{\boolean{articletitles}}{\emph{{Role of decay in the search for double-charm baryons}}, }{}\href{https://doi.org/10.1007/s11433-019-1483-0}{Sci.\ China Phys.\ Mech.\ Astron.\  \textbf{63} (2020) 221065}, \href{http://arxiv.org/abs/1912.10253}{{\normalfont\ttfamily arXiv:1912.10253}}\relax
\mciteBstWouldAddEndPuncttrue
\mciteSetBstMidEndSepPunct{\mcitedefaultmidpunct}
{\mcitedefaultendpunct}{\mcitedefaultseppunct}\relax
\EndOfBibitem
\bibitem{Hu_2020mxk}
X.-H. Hu, R.-H. Li, and Z.-P. Xing, \ifthenelse{\boolean{articletitles}}{\emph{{A comprehensive analysis of weak transition form factors for doubly heavy baryons in the light front approach}}, }{}\href{https://doi.org/10.1140/epjc/s10052-020-7851-8}{Eur.\ Phys.\ J.\  \textbf{C80} (2020) 320}, \href{http://arxiv.org/abs/2001.06375}{{\normalfont\ttfamily arXiv:2001.06375}}\relax
\mciteBstWouldAddEndPuncttrue
\mciteSetBstMidEndSepPunct{\mcitedefaultmidpunct}
{\mcitedefaultendpunct}{\mcitedefaultseppunct}\relax
\EndOfBibitem
\bibitem{Shi_2019hbf}
Y.-J. Shi, W.~Wang, and Z.-X. Zhao, \ifthenelse{\boolean{articletitles}}{\emph{{QCD sum rules analysis of weak decays of doubly-heavy baryons}}, }{}\href{https://doi.org/10.1140/epjc/s10052-020-8096-2}{Eur.\ Phys.\ J.\  \textbf{C80} (2020) 568}, \href{http://arxiv.org/abs/1902.01092}{{\normalfont\ttfamily arXiv:1902.01092}}\relax
\mciteBstWouldAddEndPuncttrue
\mciteSetBstMidEndSepPunct{\mcitedefaultmidpunct}
{\mcitedefaultendpunct}{\mcitedefaultseppunct}\relax
\EndOfBibitem
\bibitem{Shi_2017dto}
Y.-J. Shi, W.~Wang, Y.~Xing, and J.~Xu, \ifthenelse{\boolean{articletitles}}{\emph{{Weak decays of doubly heavy baryons: multi-body decay channels}}, }{}\href{https://doi.org/10.1140/epjc/s10052-018-5532-7}{Eur.\ Phys.\ J.\  \textbf{C78} (2018) 56}, \href{http://arxiv.org/abs/1712.03830}{{\normalfont\ttfamily arXiv:1712.03830}}\relax
\mciteBstWouldAddEndPuncttrue
\mciteSetBstMidEndSepPunct{\mcitedefaultmidpunct}
{\mcitedefaultendpunct}{\mcitedefaultseppunct}\relax
\EndOfBibitem
\bibitem{Shi_2020}
Y.-J. Shi, W.~Wang, Z.-X. Zhao, and U.-G. Mei\ss{}ner, \ifthenelse{\boolean{articletitles}}{\emph{{Towards a heavy diquark effective theory for weak decays of doubly heavy baryons}}, }{}\href{https://doi.org/10.1140/epjc/s10052-020-7949-z}{Eur.\ Phys.\ J.\  \textbf{C80} (2020) 398}, \href{http://arxiv.org/abs/2002.02785}{{\normalfont\ttfamily arXiv:2002.02785}}\relax
\mciteBstWouldAddEndPuncttrue
\mciteSetBstMidEndSepPunct{\mcitedefaultmidpunct}
{\mcitedefaultendpunct}{\mcitedefaultseppunct}\relax
\EndOfBibitem
\bibitem{LHCb-DP-2008-001}
LHCb collaboration, A.~A. Alves~Jr.\ {\em et~al.}, \ifthenelse{\boolean{articletitles}}{\emph{{The \lhcb detector at the LHC}}, }{}\href{https://doi.org/10.1088/1748-0221/3/08/S08005}{JINST \textbf{3} (2008) S08005}\relax
\mciteBstWouldAddEndPuncttrue
\mciteSetBstMidEndSepPunct{\mcitedefaultmidpunct}
{\mcitedefaultendpunct}{\mcitedefaultseppunct}\relax
\EndOfBibitem
\bibitem{LHCb-DP-2014-002}
LHCb collaboration, R.~Aaij {\em et~al.}, \ifthenelse{\boolean{articletitles}}{\emph{{LHCb detector performance}}, }{}\href{https://doi.org/10.1142/S0217751X15300227}{Int.\ J.\ Mod.\ Phys.\  \textbf{A30} (2015) 1530022}, \href{http://arxiv.org/abs/1412.6352}{{\normalfont\ttfamily arXiv:1412.6352}}\relax
\mciteBstWouldAddEndPuncttrue
\mciteSetBstMidEndSepPunct{\mcitedefaultmidpunct}
{\mcitedefaultendpunct}{\mcitedefaultseppunct}\relax
\EndOfBibitem
\bibitem{LHCb-DP-2014-001}
R.~Aaij {\em et~al.}, \ifthenelse{\boolean{articletitles}}{\emph{{Performance of the LHCb Vertex Locator}}, }{}\href{https://doi.org/10.1088/1748-0221/9/09/P09007}{JINST \textbf{9} (2014) P09007}, \href{http://arxiv.org/abs/1405.7808}{{\normalfont\ttfamily arXiv:1405.7808}}\relax
\mciteBstWouldAddEndPuncttrue
\mciteSetBstMidEndSepPunct{\mcitedefaultmidpunct}
{\mcitedefaultendpunct}{\mcitedefaultseppunct}\relax
\EndOfBibitem
\bibitem{LHCb-DP-2017-001}
P.~d'Argent {\em et~al.}, \ifthenelse{\boolean{articletitles}}{\emph{{Improved performance of the LHCb Outer Tracker in LHC Run 2}}, }{}\href{https://doi.org/10.1088/1748-0221/12/11/P11016}{JINST \textbf{12} (2017) P11016}, \href{http://arxiv.org/abs/1708.00819}{{\normalfont\ttfamily arXiv:1708.00819}}\relax
\mciteBstWouldAddEndPuncttrue
\mciteSetBstMidEndSepPunct{\mcitedefaultmidpunct}
{\mcitedefaultendpunct}{\mcitedefaultseppunct}\relax
\EndOfBibitem
\bibitem{LHCb-DP-2012-003}
M.~Adinolfi {\em et~al.}, \ifthenelse{\boolean{articletitles}}{\emph{{Performance of the \lhcb RICH detector at the LHC}}, }{}\href{https://doi.org/10.1140/epjc/s10052-013-2431-9}{Eur.\ Phys.\ J.\  \textbf{C73} (2013) 2431}, \href{http://arxiv.org/abs/1211.6759}{{\normalfont\ttfamily arXiv:1211.6759}}\relax
\mciteBstWouldAddEndPuncttrue
\mciteSetBstMidEndSepPunct{\mcitedefaultmidpunct}
{\mcitedefaultendpunct}{\mcitedefaultseppunct}\relax
\EndOfBibitem
\bibitem{LHCb-DP-2012-002}
A.~A. Alves~Jr.\ {\em et~al.}, \ifthenelse{\boolean{articletitles}}{\emph{{Performance of the LHCb muon system}}, }{}\href{https://doi.org/10.1088/1748-0221/8/02/P02022}{JINST \textbf{8} (2013) P02022}, \href{http://arxiv.org/abs/1211.1346}{{\normalfont\ttfamily arXiv:1211.1346}}\relax
\mciteBstWouldAddEndPuncttrue
\mciteSetBstMidEndSepPunct{\mcitedefaultmidpunct}
{\mcitedefaultendpunct}{\mcitedefaultseppunct}\relax
\EndOfBibitem
\bibitem{LHCb-DP-2012-004}
R.~Aaij {\em et~al.}, \ifthenelse{\boolean{articletitles}}{\emph{{The \lhcb trigger and its performance in 2011}}, }{}\href{https://doi.org/10.1088/1748-0221/8/04/P04022}{JINST \textbf{8} (2013) P04022}, \href{http://arxiv.org/abs/1211.3055}{{\normalfont\ttfamily arXiv:1211.3055}}\relax
\mciteBstWouldAddEndPuncttrue
\mciteSetBstMidEndSepPunct{\mcitedefaultmidpunct}
{\mcitedefaultendpunct}{\mcitedefaultseppunct}\relax
\EndOfBibitem
\bibitem{LHCb-DP-2016-001}
R.~Aaij {\em et~al.}, \ifthenelse{\boolean{articletitles}}{\emph{{Tesla: an application for real-time data analysis in high energy physics}}, }{}\href{https://doi.org/10.1016/j.cpc.2016.07.022}{Comput.\ Phys.\ Commun.\  \textbf{208} (2016) 35}, \href{http://arxiv.org/abs/1604.05596}{{\normalfont\ttfamily arXiv:1604.05596}}\relax
\mciteBstWouldAddEndPuncttrue
\mciteSetBstMidEndSepPunct{\mcitedefaultmidpunct}
{\mcitedefaultendpunct}{\mcitedefaultseppunct}\relax
\EndOfBibitem
\bibitem{LHCb-PAPER-2012-048}
LHCb collaboration, R.~Aaij {\em et~al.}, \ifthenelse{\boolean{articletitles}}{\emph{{Measurements of the \Lb, \Xibm, and \Omegab baryon masses}}, }{}\href{https://doi.org/10.1103/PhysRevLett.110.182001}{Phys.\ Rev.\ Lett.\  \textbf{110} (2013) 182001}, \href{http://arxiv.org/abs/1302.1072}{{\normalfont\ttfamily arXiv:1302.1072}}\relax
\mciteBstWouldAddEndPuncttrue
\mciteSetBstMidEndSepPunct{\mcitedefaultmidpunct}
{\mcitedefaultendpunct}{\mcitedefaultseppunct}\relax
\EndOfBibitem
\bibitem{LHCb-PAPER-2013-011}
LHCb collaboration, R.~Aaij {\em et~al.}, \ifthenelse{\boolean{articletitles}}{\emph{{Precision measurement of \D meson mass differences}}, }{}\href{https://doi.org/10.1007/JHEP06(2013)065}{JHEP \textbf{06} (2013) 065}, \href{http://arxiv.org/abs/1304.6865}{{\normalfont\ttfamily arXiv:1304.6865}}\relax
\mciteBstWouldAddEndPuncttrue
\mciteSetBstMidEndSepPunct{\mcitedefaultmidpunct}
{\mcitedefaultendpunct}{\mcitedefaultseppunct}\relax
\EndOfBibitem
\bibitem{Sjostrand:2007gs}
T.~Sj\"{o}strand, S.~Mrenna, and P.~Skands, \ifthenelse{\boolean{articletitles}}{\emph{{A brief introduction to PYTHIA 8.1}}, }{}\href{https://doi.org/10.1016/j.cpc.2008.01.036}{Comput.\ Phys.\ Commun.\  \textbf{178} (2008) 852}, \href{http://arxiv.org/abs/0710.3820}{{\normalfont\ttfamily arXiv:0710.3820}}\relax
\mciteBstWouldAddEndPuncttrue
\mciteSetBstMidEndSepPunct{\mcitedefaultmidpunct}
{\mcitedefaultendpunct}{\mcitedefaultseppunct}\relax
\EndOfBibitem
\bibitem{Sjostrand:2006za}
T.~Sj\"{o}strand, S.~Mrenna, and P.~Skands, \ifthenelse{\boolean{articletitles}}{\emph{{PYTHIA 6.4 physics and manual}}, }{}\href{https://doi.org/10.1088/1126-6708/2006/05/026}{JHEP \textbf{05} (2006) 026}, \href{http://arxiv.org/abs/hep-ph/0603175}{{\normalfont\ttfamily arXiv:hep-ph/0603175}}\relax
\mciteBstWouldAddEndPuncttrue
\mciteSetBstMidEndSepPunct{\mcitedefaultmidpunct}
{\mcitedefaultendpunct}{\mcitedefaultseppunct}\relax
\EndOfBibitem
\bibitem{LHCb-PROC-2010-056}
I.~Belyaev {\em et~al.}, \ifthenelse{\boolean{articletitles}}{\emph{{Handling of the generation of primary events in Gauss, the LHCb simulation framework}}, }{}\href{https://doi.org/10.1088/1742-6596/331/3/032047}{J.\ Phys.\ Conf.\ Ser.\  \textbf{331} (2011) 032047}\relax
\mciteBstWouldAddEndPuncttrue
\mciteSetBstMidEndSepPunct{\mcitedefaultmidpunct}
{\mcitedefaultendpunct}{\mcitedefaultseppunct}\relax
\EndOfBibitem
\bibitem{genxicc_2010}
C.-H. Chang, J.-X. Wang, and X.-G. Wu, \ifthenelse{\boolean{articletitles}}{\emph{{GENXICC2.0}: An upgraded version of the generator for hadronic production of double heavy baryons {\Xicc, \Xibc and \Xibb}}, }{}\href{https://doi.org/10.1016/j.cpc.2010.02.008}{Comput.\ Phys.\ Commun.\  \textbf{181} (2010) 1144}, \href{http://arxiv.org/abs/0910.4462}{{\normalfont\ttfamily arXiv:0910.4462}}\relax
\mciteBstWouldAddEndPuncttrue
\mciteSetBstMidEndSepPunct{\mcitedefaultmidpunct}
{\mcitedefaultendpunct}{\mcitedefaultseppunct}\relax
\EndOfBibitem
\bibitem{Lange:2001uf}
D.~J. Lange, \ifthenelse{\boolean{articletitles}}{\emph{{The EvtGen particle decay simulation package}}, }{}\href{https://doi.org/10.1016/S0168-9002(01)00089-4}{Nucl.\ Instrum.\ Meth.\  \textbf{A462} (2001) 152}\relax
\mciteBstWouldAddEndPuncttrue
\mciteSetBstMidEndSepPunct{\mcitedefaultmidpunct}
{\mcitedefaultendpunct}{\mcitedefaultseppunct}\relax
\EndOfBibitem
\bibitem{davidson2015photos}
N.~Davidson, T.~Przedzinski, and Z.~Was, \ifthenelse{\boolean{articletitles}}{\emph{{PHOTOS interface in C++: Technical and physics documentation}}, }{}\href{https://doi.org/https://doi.org/10.1016/j.cpc.2015.09.013}{Comp.\ Phys.\ Comm.\  \textbf{199} (2016) 86}, \href{http://arxiv.org/abs/1011.0937}{{\normalfont\ttfamily arXiv:1011.0937}}\relax
\mciteBstWouldAddEndPuncttrue
\mciteSetBstMidEndSepPunct{\mcitedefaultmidpunct}
{\mcitedefaultendpunct}{\mcitedefaultseppunct}\relax
\EndOfBibitem
\bibitem{Allison:2006ve}
Geant4 collaboration, J.~Allison {\em et~al.}, \ifthenelse{\boolean{articletitles}}{\emph{{Geant4 developments and applications}}, }{}\href{https://doi.org/10.1109/TNS.2006.869826}{IEEE Trans.\ Nucl.\ Sci.\  \textbf{53} (2006) 270}\relax
\mciteBstWouldAddEndPuncttrue
\mciteSetBstMidEndSepPunct{\mcitedefaultmidpunct}
{\mcitedefaultendpunct}{\mcitedefaultseppunct}\relax
\EndOfBibitem
\bibitem{Agostinelli:2002hh}
Geant4 collaboration, S.~Agostinelli {\em et~al.}, \ifthenelse{\boolean{articletitles}}{\emph{{Geant4: A simulation toolkit}}, }{}\href{https://doi.org/10.1016/S0168-9002(03)01368-8}{Nucl.\ Instrum.\ Meth.\  \textbf{A506} (2003) 250}\relax
\mciteBstWouldAddEndPuncttrue
\mciteSetBstMidEndSepPunct{\mcitedefaultmidpunct}
{\mcitedefaultendpunct}{\mcitedefaultseppunct}\relax
\EndOfBibitem
\bibitem{LHCb-PROC-2011-006}
M.~Clemencic {\em et~al.}, \ifthenelse{\boolean{articletitles}}{\emph{{The \lhcb simulation application, Gauss: Design, evolution and experience}}, }{}\href{https://doi.org/10.1088/1742-6596/331/3/032023}{J.\ Phys.\ Conf.\ Ser.\  \textbf{331} (2011) 032023}\relax
\mciteBstWouldAddEndPuncttrue
\mciteSetBstMidEndSepPunct{\mcitedefaultmidpunct}
{\mcitedefaultendpunct}{\mcitedefaultseppunct}\relax
\EndOfBibitem
\bibitem{Breiman}
L.~Breiman, J.~H. Friedman, R.~A. Olshen, and C.~J. Stone, {\em Classification and regression trees}, Wadsworth international group, Belmont, California, USA, 1984\relax
\mciteBstWouldAddEndPuncttrue
\mciteSetBstMidEndSepPunct{\mcitedefaultmidpunct}
{\mcitedefaultendpunct}{\mcitedefaultseppunct}\relax
\EndOfBibitem
\bibitem{AdaBoost}
Y.~Freund and R.~E. Schapire, \ifthenelse{\boolean{articletitles}}{\emph{A decision-theoretic generalization of on-line learning and an application to boosting}, }{}\href{https://doi.org/10.1006/jcss.1997.1504}{J.\ Comput.\ Syst.\ Sci.\  \textbf{55} (1997) 119}\relax
\mciteBstWouldAddEndPuncttrue
\mciteSetBstMidEndSepPunct{\mcitedefaultmidpunct}
{\mcitedefaultendpunct}{\mcitedefaultseppunct}\relax
\EndOfBibitem
\bibitem{Hocker:2007ht}
H.~Voss, A.~Hoecker, J.~Stelzer, and F.~Tegenfeldt, \ifthenelse{\boolean{articletitles}}{\emph{{TMVA - Toolkit for Multivariate Data Analysis with ROOT}}, }{}\href{https://doi.org/10.22323/1.050.0040}{PoS \textbf{ACAT} (2007) 040}\relax
\mciteBstWouldAddEndPuncttrue
\mciteSetBstMidEndSepPunct{\mcitedefaultmidpunct}
{\mcitedefaultendpunct}{\mcitedefaultseppunct}\relax
\EndOfBibitem
\bibitem{TMVA4}
A.~Hoecker {\em et~al.}, \ifthenelse{\boolean{articletitles}}{\emph{{TMVA 4 --- Toolkit for Multivariate Data Analysis with ROOT. Users Guide.}}, }{}\href{http://arxiv.org/abs/physics/0703039}{{\normalfont\ttfamily arXiv:physics/0703039}}\relax
\mciteBstWouldAddEndPuncttrue
\mciteSetBstMidEndSepPunct{\mcitedefaultmidpunct}
{\mcitedefaultendpunct}{\mcitedefaultseppunct}\relax
\EndOfBibitem
\bibitem{LHCb-PUB-2016-021}
L.~Anderlini {\em et~al.}, \ifthenelse{\boolean{articletitles}}{\emph{{The PIDCalib package}}, }{} \href{http://cdsweb.cern.ch/search?p=LHCb-PUB-2016-021&f=reportnumber&action_search=Search&c=LHCb+Notes} {LHCb-PUB-2016-021}, 2016\relax
\mciteBstWouldAddEndPuncttrue
\mciteSetBstMidEndSepPunct{\mcitedefaultmidpunct}
{\mcitedefaultendpunct}{\mcitedefaultseppunct}\relax
\EndOfBibitem
\bibitem{Punzi:2003bu}
G.~Punzi, \ifthenelse{\boolean{articletitles}}{\emph{{Sensitivity of searches for new signals and its optimization}}, }{}eConf \textbf{C030908} (2003) MODT002, \href{http://arxiv.org/abs/physics/0308063}{{\normalfont\ttfamily arXiv:physics/0308063}}\relax
\mciteBstWouldAddEndPuncttrue
\mciteSetBstMidEndSepPunct{\mcitedefaultmidpunct}
{\mcitedefaultendpunct}{\mcitedefaultseppunct}\relax
\EndOfBibitem
\bibitem{Skwarnicki:1986xj}
T.~Skwarnicki, {\em {A study of the radiative cascade transitions between the Upsilon-prime and Upsilon resonances}}, PhD thesis, Institute of Nuclear Physics, Krakow, 1986, {\href{http://inspirehep.net/record/230779/}{DESY-F31-86-02}}\relax
\mciteBstWouldAddEndPuncttrue
\mciteSetBstMidEndSepPunct{\mcitedefaultmidpunct}
{\mcitedefaultendpunct}{\mcitedefaultseppunct}\relax
\EndOfBibitem
\bibitem{Cowan:2016tnm}
G.~A. Cowan, D.~C. Craik, and M.~D. Needham, \ifthenelse{\boolean{articletitles}}{\emph{{RapidSim: an application for the fast simulation of heavy-quark hadron decays}}, }{}\href{https://doi.org/10.1016/j.cpc.2017.01.029}{Comput.\ Phys.\ Commun.\  \textbf{214} (2017) 239}, \href{http://arxiv.org/abs/1612.07489}{{\normalfont\ttfamily arXiv:1612.07489}}\relax
\mciteBstWouldAddEndPuncttrue
\mciteSetBstMidEndSepPunct{\mcitedefaultmidpunct}
{\mcitedefaultendpunct}{\mcitedefaultseppunct}\relax
\EndOfBibitem
\bibitem{significance_2011}
G.~Cowan, K.~Cranmer, E.~Gross, and O.~Vitells, \ifthenelse{\boolean{articletitles}}{\emph{Asymptotic formulae for likelihood-based tests of new physics}, }{}\href{https://doi.org/10.1140/epjc/s10052-011-1554-0}{Eur.\ Phys.\ J.\  \textbf{C71} (2011) 1554}, \href{http://arxiv.org/abs/1007.1727}{{\normalfont\ttfamily arXiv:1007.1727}}\relax
\mciteBstWouldAddEndPuncttrue
\mciteSetBstMidEndSepPunct{\mcitedefaultmidpunct}
{\mcitedefaultendpunct}{\mcitedefaultseppunct}\relax
\EndOfBibitem
\bibitem{Pivk:2004ty}
M.~Pivk and F.~R. Le~Diberder, \ifthenelse{\boolean{articletitles}}{\emph{{sPlot: A statistical tool to unfold data distributions}}, }{}\href{https://doi.org/10.1016/j.nima.2005.08.106}{Nucl.\ Instrum.\ Meth.\  \textbf{A555} (2005) 356}, \href{http://arxiv.org/abs/physics/0402083}{{\normalfont\ttfamily arXiv:physics/0402083}}\relax
\mciteBstWouldAddEndPuncttrue
\mciteSetBstMidEndSepPunct{\mcitedefaultmidpunct}
{\mcitedefaultendpunct}{\mcitedefaultseppunct}\relax
\EndOfBibitem
\bibitem{LHCb-DP-2013-002}
LHCb collaboration, R.~Aaij {\em et~al.}, \ifthenelse{\boolean{articletitles}}{\emph{{Measurement of the track reconstruction efficiency at LHCb}}, }{}\href{https://doi.org/10.1088/1748-0221/10/02/P02007}{JINST \textbf{10} (2015) P02007}, \href{http://arxiv.org/abs/1408.1251}{{\normalfont\ttfamily arXiv:1408.1251}}\relax
\mciteBstWouldAddEndPuncttrue
\mciteSetBstMidEndSepPunct{\mcitedefaultmidpunct}
{\mcitedefaultendpunct}{\mcitedefaultseppunct}\relax
\EndOfBibitem
\bibitem{Rogozhnikov:2016bdp}
A.~Rogozhnikov, \ifthenelse{\boolean{articletitles}}{\emph{{Reweighting with boosted decision trees}}, }{}\href{https://doi.org/10.1088/1742-6596/762/1/012036}{J.\ Phys.\ Conf.\ Ser.\  \textbf{762} (2016) 012036}, \href{http://arxiv.org/abs/1608.05806}{{\normalfont\ttfamily arXiv:1608.05806}}, \url{https://github.com/arogozhnikov/hep_ml}\relax
\mciteBstWouldAddEndPuncttrue
\mciteSetBstMidEndSepPunct{\mcitedefaultmidpunct}
{\mcitedefaultendpunct}{\mcitedefaultseppunct}\relax
\EndOfBibitem
\bibitem{LHCb-PAPER-2016-008}
LHCb collaboration, R.~Aaij {\em et~al.}, \ifthenelse{\boolean{articletitles}}{\emph{{Measurements of the mass and lifetime of the \Omegab baryon}}, }{}\href{https://doi.org/10.1103/PhysRevD.93.092007}{Phys.\ Rev.\  \textbf{D93} (2016) 092007}, \href{http://arxiv.org/abs/1604.01412}{{\normalfont\ttfamily arXiv:1604.01412}}\relax
\mciteBstWouldAddEndPuncttrue
\mciteSetBstMidEndSepPunct{\mcitedefaultmidpunct}
{\mcitedefaultendpunct}{\mcitedefaultseppunct}\relax
\EndOfBibitem
\bibitem{Lindenbaum:1957ec}
S.~J. Lindenbaum and R.~M. Sternheimer, \ifthenelse{\boolean{articletitles}}{\emph{{Isobaric nucleon model for pion production in nucleon-nucleon collisions}}, }{}\href{https://doi.org/10.1103/PhysRev.105.1874}{Phys.\ Rev.\  \textbf{105} (1957) 1874}\relax
\mciteBstWouldAddEndPuncttrue
\mciteSetBstMidEndSepPunct{\mcitedefaultmidpunct}
{\mcitedefaultendpunct}{\mcitedefaultseppunct}\relax
\EndOfBibitem
\bibitem{LHCb-DP-2020-001}
C.~Abellan~Beteta {\em et~al.}, \ifthenelse{\boolean{articletitles}}{\emph{{Calibration and performance of the LHCb calorimeters in Run 1 and 2 at the LHC}}, }{}\href{http://arxiv.org/abs/2008.11556}{{\normalfont\ttfamily arXiv:2008.11556}}, {submitted to JINST}\relax
\mciteBstWouldAddEndPuncttrue
\mciteSetBstMidEndSepPunct{\mcitedefaultmidpunct}
{\mcitedefaultendpunct}{\mcitedefaultseppunct}\relax
\EndOfBibitem
\bibitem{Nisius:2014wua}
R.~Nisius, \ifthenelse{\boolean{articletitles}}{\emph{{On the combination of correlated estimates of a physics observable}}, }{}\href{https://doi.org/10.1140/epjc/s10052-014-3004-2}{Eur.\ Phys.\ J.\  \textbf{C74} (2014) 3004}, \href{http://arxiv.org/abs/1402.4016}{{\normalfont\ttfamily arXiv:1402.4016}}\relax
\mciteBstWouldAddEndPuncttrue
\mciteSetBstMidEndSepPunct{\mcitedefaultmidpunct}
{\mcitedefaultendpunct}{\mcitedefaultseppunct}\relax
\EndOfBibitem
\bibitem{Nisius:2020jmf}
R.~Nisius, \ifthenelse{\boolean{articletitles}}{\emph{{BLUE: combining correlated estimates of physics observables within ROOT using the Best Linear Unbiased Estimate method}}, }{}\href{https://doi.org/10.1016/j.softx.2020.100468}{SoftwareX \textbf{11} (2020) 100468}, \href{http://arxiv.org/abs/2001.10310}{{\normalfont\ttfamily arXiv:2001.10310}}\relax
\mciteBstWouldAddEndPuncttrue
\mciteSetBstMidEndSepPunct{\mcitedefaultmidpunct}
{\mcitedefaultendpunct}{\mcitedefaultseppunct}\relax
\EndOfBibitem
\bibitem{Wang_2017mqp}
W.~Wang, F.-S. Yu, and Z.-X. Zhao, \ifthenelse{\boolean{articletitles}}{\emph{{Weak decays of doubly heavy baryons: the $1/2\rightarrow 1/2$ case}}, }{}\href{https://doi.org/10.1140/epjc/s10052-017-5360-1}{Eur.\ Phys.\ J.\  \textbf{C77} (2017) 781}, \href{http://arxiv.org/abs/1707.02834}{{\normalfont\ttfamily arXiv:1707.02834}}\relax
\mciteBstWouldAddEndPuncttrue
\mciteSetBstMidEndSepPunct{\mcitedefaultmidpunct}
{\mcitedefaultendpunct}{\mcitedefaultseppunct}\relax
\EndOfBibitem
\bibitem{Li_2021rfj}
D.-M. Li, X.-R. Zhang, Y.~Xing, and J.~Xu, \ifthenelse{\boolean{articletitles}}{\emph{{Weak decays of doubly heavy baryons: four-body nonleptonic decay channels}}, }{}\href{https://doi.org/10.1140/epjp/s13360-021-01757-6}{Eur.\ Phys.\ J.\ Plus \textbf{136} (2021) 772}, \href{http://arxiv.org/abs/2101.12574}{{\normalfont\ttfamily arXiv:2101.12574}}\relax
\mciteBstWouldAddEndPuncttrue
\mciteSetBstMidEndSepPunct{\mcitedefaultmidpunct}
{\mcitedefaultendpunct}{\mcitedefaultseppunct}\relax
\EndOfBibitem
\end{mcitethebibliography}
\newpage
\centerline
{\large\bf LHCb collaboration}
\begin
{flushleft}
\small
R.~Aaij$^{38}$\lhcborcid{0000-0003-0533-1952},
A.S.W.~Abdelmotteleb$^{57}$\lhcborcid{0000-0001-7905-0542},
C.~Abellan~Beteta$^{51}$\lhcborcid{0009-0009-0869-6798},
F.~Abudin{\'e}n$^{57}$\lhcborcid{0000-0002-6737-3528},
T.~Ackernley$^{61}$\lhcborcid{0000-0002-5951-3498},
A. A. ~Adefisoye$^{69}$\lhcborcid{0000-0003-2448-1550},
B.~Adeva$^{47}$\lhcborcid{0000-0001-9756-3712},
M.~Adinolfi$^{55}$\lhcborcid{0000-0002-1326-1264},
P.~Adlarson$^{83}$\lhcborcid{0000-0001-6280-3851},
C.~Agapopoulou$^{14}$\lhcborcid{0000-0002-2368-0147}
% ...

{\footnotesize \it

% ...
$^{14}$Universit{\'e} Paris-Saclay, CNRS/IN2P3, IJCLab, Orsay, France\\
% ...
$^{38}$Nikhef National Institute for Subatomic Physics, Amsterdam, Netherlands\\
% ...
$^{47}$Instituto Galego de F{\'\i}sica de Altas Enerx{\'\i}as (IGFAE), Universidade de Santiago de Compostela, Santiago de Compostela, Spain\\
% ...
$^{51}$Physik-Institut, Universit{\"a}t Z{\"u}rich, Z{\"u}rich, Switzerland\\
% ...
$^{55}$H.H. Wills Physics Laboratory, University of Bristol, Bristol, United Kingdom\\
% ...
$^{57}$Department of Physics, University of Warwick, Coventry, United Kingdom\\
% ...
$^{60}$School of Physics and Astronomy, University of Glasgow, Glasgow, United Kingdom\\
$^{61}$Oliver Lodge Laboratory, University of Liverpool, Liverpool, United Kingdom\\
% ...
$^{69}$Syracuse University, Syracuse, NY, United States\\
% ...
$^{83}$Department of Physics and Astronomy, Uppsala University, Uppsala, Sweden, associated to $^{60}$\\
% ...
}
\end{flushleft}

\end{document}